\documentclass[12pt,a4paper,superscriptaddress,showkeys,email]{article}

\usepackage{epsfig} 
\usepackage{amsmath} 
\usepackage{amssymb} 
\usepackage{bbm} 
\usepackage{wasysym} 
\usepackage{mathrsfs} 
\usepackage{graphicx} 
\usepackage{amsfonts} 
\usepackage{amsbsy} 
\usepackage{amscd} 
\usepackage{bbm}  
\usepackage{accents} 
\usepackage{axodraw} 
\usepackage{pstricks} 
\usepackage{subfigure}
\usepackage{floatflt}
\newcommand{\beq}{\begin{equation}} 
\newcommand{\eeq}{\end{equation}} 
\newcommand{\beqa}{\begin{eqnarray}} 
\newcommand{\eeqa}{\end{eqnarray}} 
\newcommand{\barr}{\begin{array}} 
\newcommand{\earr}{\end{array}} 
\newcommand{\bit}{\begin{itemize}} 
\newcommand{\eit}{\end{itemize}} 
 
\newcommand{\nn}{\nonumber} 
\newcommand{\Sb}{\overline{\Sigma}} 
\newcommand{\dmsq}{\Delta m^2} 
\newcommand{\dma}{{\Delta m}^2_{\rm atm}} 
\newcommand{\dms}{{\Delta m}^2_\odot} 
\newcommand{\D}{D\hspace{-8pt}\slash}

\newcommand{\del}{\partial\hspace{-6pt}\slash} 

\newcommand{\calG}{\mathcal{G}}

\newcommand{\Qn}{\accentset{(n)}{Q}}

\DeclareMathOperator{\re}{Re} 
\DeclareMathOperator{\im}{Im} 
 
\DeclareMathOperator{\Tr}{Tr} 
 
\begin{document}

\author{Shamayita Ray\footnote{
Electronic address:~sr643@cornell.edu
\newline 
$\mbox{\;  \; \;}$Alternative electronic address:~shamayitar@theory.tifr.res.in}
\\
{\it Institute for High Energy Phenomenology,}\\
{\it Newman Laboratory of Elementary Particle Physics,}\\
{\it Cornell University, Ithaca, NY 14853, USA} }

\title{Renormalization group evolution of neutrino masses and mixing in seesaw models: A review}

\date{\today}


\maketitle

\begin{abstract}

We consider different extensions of the standard model 
which can give rise to the small active neutrino masses through seesaw 
mechanisms, and their mixing. These tiny neutrino masses are 
generated at some high energy scale by the heavy seesaw fields which 
then get sequentially decoupled to give an effective dimension-5 
operator. The renormalization group evolution of the 
masses and the mixing parameters of the three active neutrinos 
in the high energy as well as the low energy effective theory is 
reviewed in this article.

\end{abstract}

\newpage
\tableofcontents
\setcounter{tocdepth}{3}



\section{Introduction}
\label{sec:intro}

\subsection{Neutrino oscillations: The current status}

The field of neutrino physics has made immense progress in the 
last decade, which was initiated when the Super-Kamiokande (SK) 
experiment in Japan \cite{SK} reported the evidence for 
oscillations in the atmospheric neutrinos. Now 
there is compelling evidence that solar, atmospheric, 
accelerator and reactor neutrinos oscillate, which implies 
that the neutrinos are massive and the leptons mix among themselves.

\begin{figure}[h!]
\begin{center}
\parbox[l]{6.0cm}{
\subfigure[]{
\includegraphics[width=5.5cm]{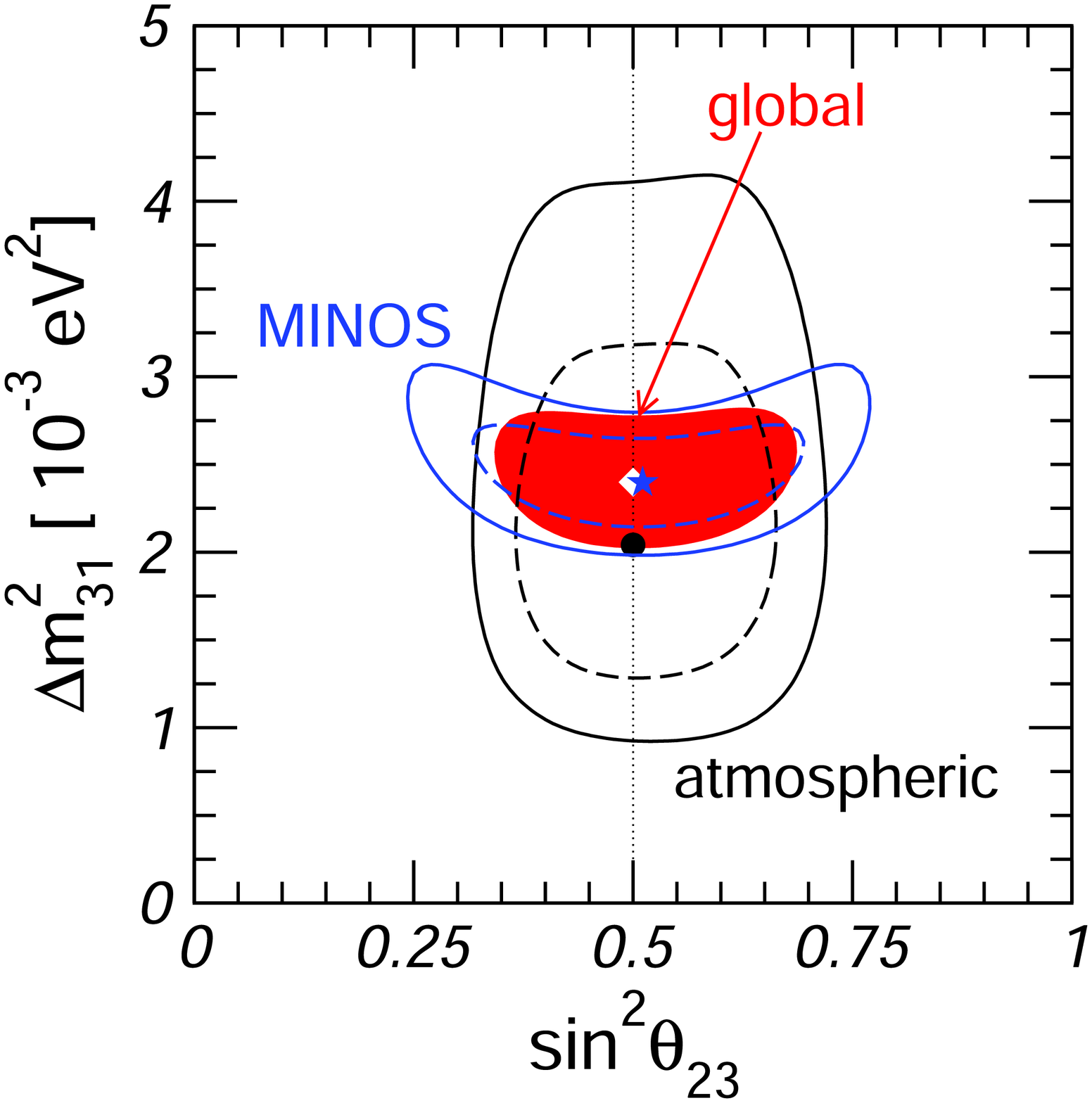}
\label{atm-fit}}
}
\parbox[r]{6.0cm}{
\subfigure[]{
\includegraphics[height=5.75cm,angle=-90]{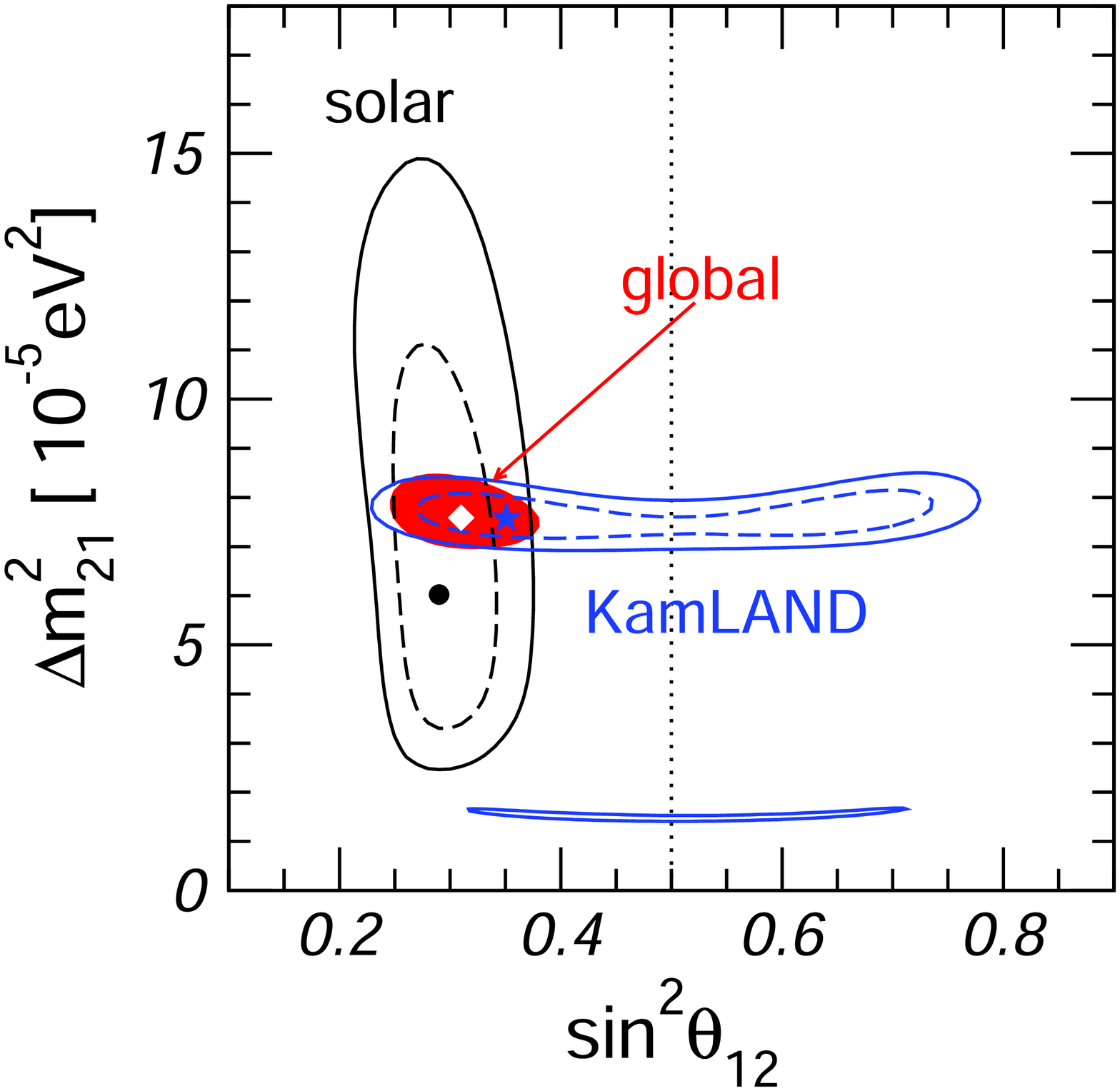}
\label{solar-fit}}
}
\caption{(a) The region of the atmospheric oscillation parameters 
$\dma$ and $\sin^2{2\theta_{\rm atm}}$ obtained from the SK, K2K 
and MINOS experiments \cite{Schwetz}; $\quad$ 
(b) The allowed region in the 
neutrino oscillation parameter space from solar neutrino data 
and KamLAND experiment \cite{Schwetz}.}
\end{center}
\end{figure}

The atmospheric neutrinos are produced in the Earth's atmosphere 
by cosmic rays. The flux of cosmic rays that lead to 
neutrinos with energies above a few GeV is isotropic. 
Hence one expects the downward and the upward-going fluxes 
of multi-GeV neutrinos of a given flavor to be equal. 
The underground SK detector found that for multi-GeV
atmospheric muon neutrinos the zenith-angle dependence deviates 
from this expectation and the deviation can be explained 
when one invokes $\nu_\mu \to \nu_\tau$ oscillations. 
The oscillations of muon neutrinos into other flavors have also been 
confirmed by the energy spectrum obtained from the controlled 
source experiments K2K \cite{k2k:2006} and MINOS \cite{minos}. 
The allowed region for the oscillation parameters, $\dma$ and 
$\sin^2{2\theta_{\rm atm}}$, is shown in Fig~\ref{atm-fit}.
As can be seen from the figure, the MINOS data 
is capable of measuring $\dma$ with high precision, while 
SK put stronger bound on $\sin^2{2\theta_{\rm atm}}$.
The results from the short-baseline (SBL) experiments 
(like CDHS \cite{Dydak:1983zq}, NOMAD \cite{Astier:2003gs} etc.) 
show that the $\nu_\mu \to \nu_e$ oscillations can be 
present only as small sub-dominant effects and also 
put strong bounds on the active-sterile mixing angles in 
$\nu_\mu \to \nu_s$ oscillations, an oscillation channel 
whose sub-dominant effect is not yet ruled out completely.

The pioneering solar neutrino experiment by Davis and collaborators 
using $^{37}$Cl reported a solar electron neutrino flux significantly smaller 
than that predicted by the standard solar model, and this deficit 
in the number of electron neutrinos is known as the ``solar neutrino problem''. 
The puzzle persisted in the literature for about 30 years, and then 
the charged current (CC) and the neutral current (NC) data from the 
SNO experiment \cite{SNO}, combined with the SK solar neutrino 
data \cite{SK-solar}, provided direct evidence for neutrino 
oscillations in solar neutrinos. However, four different solutions 
were there to explain the solar neutrino oscillations \cite{Bahcall:2002hv}: 
(i) the LMA or large mixing angle solution 
($\dms = 5.0 \times 10^{-5}$ eV$^2, \; \tan^2\theta_\odot=0.42$), 
(ii) the low mass solution ($\dms = 7.9 \times 10^{-8}$ eV$^2, \; \tan^2\theta_\odot=0.61$), 
(iii) the vacuum solution ($\dms = 4.6 \times 10^{-10}$ eV$^2, \; \tan^2\theta_\odot=1.8$) and 
(iv) the SMA or small mixing angle solution 
($\dms = 5.0 \times 10^{-6}$ eV$^2, \; \tan^2\theta_\odot= 1.5.10^{-3}$). 
The results from the controlled source experiment KamLAND \cite{Kamland} 
confirmed the LMA solution and ruled out the 
other three possibilities. Fig~\ref{solar-fit} shows the allowed region of the 
solar neutrino oscillation parameters $\dms$ and $\tan^2\theta_\odot$.

Combining the results obtained from the solar, atmospheric and 
the reactor neutrino oscillation experiments described above, 
the current knowledge about the neutrinos is that there are 
three neutrino flavors ($\nu_\alpha, \alpha \in
\{e,\mu,\tau\}$) which mix to form three neutrino mass eigenstates
($\nu_i, i \in \{1,2,3\}$). These mass eigenstates are separated by
$\dmsq_{ij} \equiv m_i^2 - m_j^2$ where, $m_{i,j}$ denote mass 
eigenvalues  with $ i,j \in \{ 1,2,3 \}$.
The two sets of eigenstates are connected through 
$\nu_\alpha = (U_{\rm PMNS})_{\alpha i} \nu_i$,
where $U_{\rm PMNS}$ is the Pontecorvo-Maki-Nakagawa-Sakata 
neutrino mixing matrix 
\cite{pontecorvo-1,pontecorvo-2, pontecorvo-gribov, mns} 
in the basis where the charged lepton mass matrix is diagonal. 
This mixing matrix is parametrized as
\beq
U_{\rm PMNS} = P \cdot {\cal U} \cdot Q \; , 
\label{Upmns}
\eeq
where 
\beqa
{\cal U} &=& U_{23}(\theta_{23},0) \; U_{13}(\theta_{13},\delta) \; 
U_{12}(\theta_{12},0) \; , \quad 
Q = {\rm Diag}\{e^{-i \phi_1},e^{-i \phi_2},1 \} \;. \quad
\label{Uckm}
\eeqa
Here $U_{ij}(\theta, \delta)$ is the complex rotation matrix 
in the $i$-$j$ plane, $\delta$ is the 
Dirac CP violating phase, $\phi_i$ are the Majorana phases, 
and $P$ is the flavor phase matrix (Sometimes the flavor phases are called as the 
unphysical phases since they do not
play any role in the phenomenology of neutrino mixing or beta-decay.) 
Finally, with all the above definitions, ${\cal U}$ takes the form
\beqa 
{\cal U} = \left( 
 \barr{ccc} 
 c_{12}c_{13} & s_{12}c_{13} & s_{13}e^{-i \delta}\\ 
 -c_{23}s_{12}-s_{23}s_{13}c_{12}e^{i \delta} & 
 c_{23}c_{12}-s_{23}s_{13}s_{12}e^{i \delta} & s_{23}c_{13}\\ 
 s_{23}s_{12}-c_{23}s_{13}c_{12}e^{i \delta} & 
 -s_{23}c_{12}-c_{23}s_{13}s_{12}e^{i \delta} & c_{23}c_{13} 
 \earr 
 \right) \; , 
\label{U}
\eeqa 
where $c_{ij}$ and $s_{ij}$ are the cosines and sines respectively 
of the mixing angle $\theta_{ij}$.
The current best-fit values and 3$\sigma$ ranges of these 
parameters are summarized in Table~\ref{tab:bounds}. 
It is still not known whether the neutrino mass ordering 
is normal ($m_1 < m_2 < m_3$) 
or inverted ($m_3 < m_1 < m_2$).
\begin{table}[t]
\begin{center}
\begin{tabular}{ccccc}\hline
& \phantom{space} & Best fit & \phantom{space} & $3\sigma$ range   \\\hline
$\Delta m_{21}^2$ [$10^{-5}{\rm eV}^2$] & & 7.65 & & 7.05 - 8.34 \\
$|\Delta m_{31}^2|$ [$10^{-3}{\rm eV}^2$] & & 2.40 & & 2.07 - 2.75 \\\hline
$\sin^2\theta_{12}$ && 0.304 && 0.25 - 0.37 \\
$\sin^2\theta_{23}$ && 0.50 && 0.36 - 0.67 \\
$\sin^2\theta_{13}$ && 0.01 && $\leq$ 0.056 \\\hline
\end{tabular}
\caption{The present best-fit values and 3$\sigma$ ranges
of oscillation parameters \cite{Schwetz, Fogli, Goswami}.
\label{tab:bounds}}
\end{center}
\end{table}
Many other high precision oscillation experiments are going on and 
also being planned in order to measure the neutrino oscillation parameters 
with higher accuracy and to determine the neutrino mass ordering.

As can be seen from the PMNS parametrization of the neutrino mixing matrix in Eq.~(\ref{U}), 
the angle $\theta_{13}$ plays a crucial role in the determination of the Dirac 
CP phase $\delta$. As shown in the Table~\ref{tab:bounds}, $\theta_{13}$ can 
also be consistent with zero at 3$\sigma$. However, this data also implies that assuming 
the error to scale linearly upto 3$\sigma$ within the 
physical range of $\sin^2{\theta_{13}}$, there is a hint of $\theta_{13} > 0$ 
at $\sim$0.9$\sigma$. It has been shown that the solar and KamLAND data 
implies a non-zero $\theta_{13}$ at $\sim$1.5$\sigma$ 
\cite{Schwetz-th13-nonzero,Lisi-th13-nonzero}. But when combined 
with atmospheric, long-baseline reactor and CHOOZ data, the 
significance is lowered since the hint for a non-zero $\theta_{13}$ 
from the atmospheric data is not so robust and depends on the 
details of event rate calculations and the treatment of theoretical 
uncertainties \cite{Schwetz-th13-nonzero}.


\subsection{Absolute masses of the active neutrinos}

While the neutrino oscillation experiments are not sensitive to the 
absolute neutrino masses, the beta decay and the neutrinoless 
double beta decay ($0\nu \beta \beta$) processes are. 
At the same time, it is possible to estimate $\sum_i m_i$ 
from cosmology also. 
In case of beta decay, the non-zero neutrino mass 
would modify the Kurie plot, regardless of whether the neutrinos 
are Dirac or Majorana particles. The effect will depend on 
$m_\beta = {\left( \sum_i |U_{ei}|^2 m_i^2 \right)}^{1/2}$, 
and if the neutrino masses are small, it will be visible 
only near the end point of the Kurie plot.
The Mainz \cite{mainz} experiment has placed the upper limit of 
$m_\beta \leq 2.3$ eV. The upcoming beta-decay 
experiments like KATRIN \cite{katrin} will be sensitive to 
$m_\beta > 0.2$ eV and will thus improve the bound by an 
order of magnitude. 
The $0\nu \beta \beta$ decay, 
on the other hand, is sensitive to the effective Majorana mass of 
the electron neutrinos, defined as $m_{ee} \equiv |\sum_i U_{ei}^2 m_i|$, 
and will be observed only if the neutrinos are Majorana particles. 
A non-zero signal for the $0\nu \beta \beta$ decay will 
put bound on the specific combination of the neutrino masses and 
the Majorana phases given by $m_{ee}$. 
The current limit put by the Heidelberg-Moscow 
experiment \cite{heidelberg-moscow} is $m_{ee} \lesssim 0.9$ eV. 
The cosmic microwave background radiation (CMBR) carries 
the imprint of the neutrino masses since in the standard 
Big Bang model, for the standard model (SM) interactions 
of the neutrinos, the neutrinos are abundant like the photons till the 
epoch of nucleosynthesis when they decouple from the thermal 
bath of the photons. It is also possible to get information about the 
neutrino masses from the study of the large scale structure as an active 
neutrino species of mass $m_\nu$ will tend to wash out all structures 
upto a scale $\sim 1/m_\nu$ by free-streaming. Recent results 
from the Wilkinson Microwave Anisotropy Probe (WMAP) and the 
surveys on the large scale structure put the limit 
$\sum_i m_i \leq $ 0.67 - 2.0 eV \cite{WMAP5,Hannestad}.

The very fact that the active neutrinos are massive demands an extension of the 
SM. In the framework of the SM, since there is no right-handed neutrino, 
the neutrinos are massless at the tree-level, and they 
cannot have a Dirac mass even at loop level.
So the only other possibility is the lepton number violating Majorana 
mass term. But lepton number is a symmetry of the SM, though accidental, 
and if that symmetry is to be obeyed, Majorana masses also cannot be 
generated at loop level. 
It can also be seen that the Planck scale ($M_{\rm Pl}$) effect 
cannot introduce the required neutrino mass in the SM 
as it can only generate a neutrino mass 
$\sim {\cal O}(v_{EW}^2/M_{\rm Pl}) \approx {\cal O}(10^{-5} {\rm eV})$, 
and hence cannot explain the atmospheric mass squared difference.
Hence generally the neutrino masses are incorporated at the tree-level by 
adding new fields to the SM at high energy scales.
The most favored mechanisms to 
generate such small neutrino masses are the so called seesaw mechanisms which
need the introduction of one or more heavy fields, while maintains the 
SU(3)$_C \times$ SU(2)$_L \times$ U(1)$_Y$ gauge group structure of the SM.
There are also other models like Inverse seesaw \cite{Mohapatra:1986bd}, 
the model with a singly charged singlet proposed originally by Zee 
\cite{zee-model,Pal-Mohapatra}, the model with a doubly charged singlet 
\cite{Pal-Mohapatra,doubly-charged-scalar}, etc. 
Recently another new model has been proposed in \cite{babu-mass}, where 
a pair of vector like leptons and also a Higgs quadruplet are added 
to the SM to generate neutrino mass. However, some models, like the Zee's 
model, cannot predict neutrino mixing parameters consistent with 
the current data. We will discuss some of these models of neutrino 
masses in detail in Section~\ref{sec:mass-generation}.


\subsection{RG evolution of neutrino parameters}

Since the neutrino mass is generated at the high scale 
while the neutrino masses and mixing parameters are measured experimentally at a low
scale, the renormalization group (RG) evolution effects need to be included. 
The current experimental data in Table~\ref{tab:bounds} shows that 
in the neutrino sector two of the three mixing angles are 
large, while the third one is small, which is rather 
different from the quark sector where all three mixing angles are small. 
Because of the large values of the two mixing angles, RG
evolution of the neutrino masses and the mixing parameters 
plays an important role in the neutrino sector, which is not the case with the quark sector.
RG evolution will be even larger if the neutrinos happen to be quasi-degenerate. 

The radiative corrections in different theories of neutrino masses are 
expected to be different since the heavy particles couple differently 
to the SM fields present. However below the mass scale of the lightest 
of the heavy particles the effect of all heavy degrees of freedom are 
integrated out to get an effective theory of neutrino masses. The 
RG evolution of neutrino masses and mixing parameters in the low energy 
effective theory as well as in different high energy theories will be 
discussed in Section~\ref{sec:RG_effective} and Section~\ref{sec:RG_full} 
respectively.


\section{Generation of light neutrino masses}
\label{sec:mass-generation}

\subsection{Low energy effective theory of neutrino masses}
\label{subsec:effective_L}

The low energy effective Lagrangian needed to explain 
the non-zero active neutrino masses can 
in general be expressed as a series of non-renormalizable operators, the 
dominant one being the dimension-5 operator given as \cite{weinberg}
\beq
{\cal L}_\kappa \sim \kappa_5 l_L l_L \phi \phi \; .
\label{dim-5 L}
\eeq
where $l_L$ and $\phi$ are respectively the lepton and Higgs doublets belonging to the SM. 
Here $\kappa_5$ is the effective coupling which 
can be expressed in terms of a dimensionless coupling $a_5$ as 
$\kappa_5 = a_5 /\Lambda$ with $\Lambda$ some high energy scale. 
In this picture the SM serves as an effective theory 
valid upto the mass scale $\Lambda$, which can be taken to be the mass of the lightest of 
the heavy fields. 
\begin{figure} 
\begin{center} 
\includegraphics[scale=1.2]{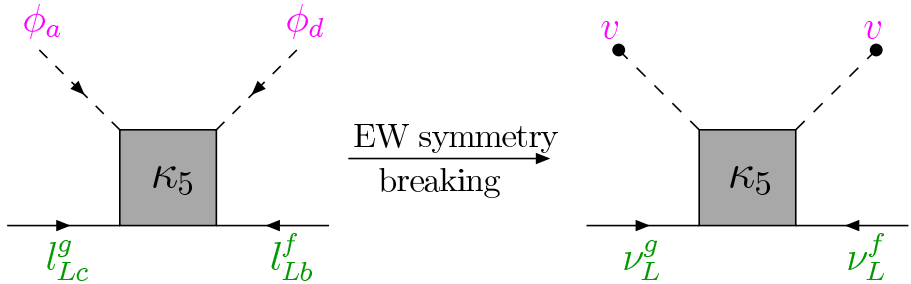}
\caption{Generation of the Majorana neutrino mass from the low energy 
effective Lagrangian given in Eq.~(\ref{dim-5 L}) after spontaneous 
symmetry breaking.
\label{effective-low-energy}} 
\end{center} 
\end{figure} 
However, the specific form of $\kappa_5$ will depend on the high 
energy field content and the interactions present at the high scale. 
The operator shown in Eq.~(\ref{dim-5 L}) violates lepton number by two units 
and gives rise to Majorana masses for neutrinos after spontaneous 
symmetry breaking, ${\mathbbm m}_\nu \sim \frac{1}{2}\kappa_5 v^2$, 
as shown in the Fig.~\ref{effective-low-energy}. Here 
$v$ is the vacuum expectation value (vev) of the Higgs field $\phi$ such that 
\beq
\phi = \left( \barr{c} \phi^+ \\ \phi^0 \earr \right) 
\xrightarrow[\text{symmetry breaking}]{\text{Spontaneous}} 
\left( \barr{c} 0 \\ \frac{v}{\sqrt{2}} \earr \right)\; . 
\label{vev}
\eeq
Taking $v \sim 246$ GeV, a neutrino mass of $\sim 0.05$ eV implies 
$\Lambda \sim 10^{15}$ GeV if $a_5 \sim 1$.


\subsection{High energy theories: Seesaw mechanisms}

There are four possible ways to form a dimension-5 gauge singlet term as 
given in Eq.~(\ref{dim-5 L}) 
at low energy through the tree-level exchange 
of a heavy particle at the high energy: 
(i) each $l_L$-$\phi$ pair forms a fermion singlet, 
(ii) each of the $l_L$-$l_L$ and $\phi$-$\phi$ pair forms a scalar triplet, 
(iii) each $l_L$-$\phi$ pair forms a fermion triplet,
and (iv) each of the $l_L$-$l_L$ and $\phi$-$\phi$ pair forms a scalar singlet. 
Case (i) can arise from the tree-level exchange of a right handed fermion singlet 
and this corresponds to the Type-I seesaw mechanism
\cite{Minkowski:1977sc,Yanagida-Type-I,GellMann-Type-I,Glashow-Type-I,Mohapatra:1979ia}. 
Case (ii) arises when the heavy particle is a Higgs triplet giving rise to the
Type-II seesaw mechanism \cite{Magg:1980ut,Lazarides:1980nt}. 
For case (iii) the exchanged particle 
should be a right-handed fermion triplet, which corresponds to 
the Type-III seesaw mechanism \cite{Foot-Type-III,Ma:2002pf}. 
The last scenario gives terms only of the form $\overline{\nu_L^C} e_L$, which 
cannot generate a neutrino mass. We describe the three different seesaw mechanisms 
in Section~\ref{sec:typeI}--\ref{sec:typeIII} in detail. A summary of the 
form of $\kappa$ and hence the effective light neutrino mass
at the low scale in different types of seesaw is given at the 
and of this section in Table~\ref{seesaw-summary}. 
There is another model, similar to the seesaw models, that can predict 
the light active neutrino masses and known as the inverse seesaw 
model. This model will be discussed in Sec~\ref{sec:inverse_seesaw}, 
for the sake of completeness.


\subsubsection{Type-I seesaw}
\label{sec:typeI}

The simplest extension of the SM to incorporate small active neutrino 
mass is to introduce right-handed singlet fermions $N_R$ in the 
theory, which are singlets under the SM gauge group. 
Hence these $N_R$ fields are essentially right-handed neutrinos. 
Presence of these new fields allows new terms in the Lagrangian
\beqa
{\cal L}_N = \frac{1}{2} \overline{N} (i \del) 
N - \frac{1}{2} \overline{N} {\mathbbm M}_{\rm N} N 
- \Bigl( \overline{N} Y_{\rm N} \widetilde{\phi}^\dagger l_L + {\text {h.c.}} \Bigr) \; , 
\label{L-N}
\eeqa
where $l_L$ and $\phi$ are respectively the lepton and Higgs doublets belonging to 
the SM and $\widetilde\phi \equiv i \sigma^2 \phi^\ast$, $\sigma^2$ being the second 
Pauli matrix. 
Here we do not write the generation or the SU(2)$_L$ indices explicitly. The field
$N$ is defined as $N \equiv N_R + N_R^C$, where $N_R^C$ is the CP conjugate of the 
right-handed field $N_R$. $Y_{\rm N}$ is the Yukawa coupling for the singlet 
fermion and ${\mathbbm M}_{\rm N}$ is the mass matrix. Thus the complete Lagrangian 
of the theory becomes
\beq
{\cal L} = {\cal L}_{\rm SM} + {\cal L}_N \; ,
\eeq
and after spontaneous symmetry breaking it is possible to write the neutrino mass terms as
\beqa
-{\cal L}_{\nu_{\rm mass}} = \frac{1}{2}
\left( \barr{cc} \overline{\nu_L} & \overline{N_R^C} \earr \right) 
\left( \barr{cc} 0 & {\mathbbm m}_D \\ {\mathbbm m}_D^T & {\mathbbm M}_{\rm N} \earr \right) 
\left( \barr{c} \nu_L^C \\ N_R \earr \right) + {\text {h.c.}} \; ,
\eeqa
where 
\beq
{\mathbbm m}_D = (v/\sqrt{2}) Y_{\rm N}^T
\eeq
is the Dirac mass matrix for the 
neutrinos generated after the electroweak symmetry breaking when the Higgs 
gets the vev $v$, as given in Eq.~(\ref{vev}). Thus the complete mass 
matrix for the neutrinos becomes
\beqa
{\cal M}_\nu = \left( \barr{cc} 0 & {\mathbbm m}_D \\ 
{\mathbbm m}_D^T & {\mathbbm M}_{\rm N} \earr \right) \; ,
\label{mass-matrix-nu}
\eeqa
which when block-diagonalized gives the eigenvalues (see Appendix for derivation)
\beqa
{\mathbbm m}_1 &\approx& -{\mathbbm m}_D {\mathbbm M}_{\rm N}^{-1} 
{\mathbbm m}_D^T \; , 
\label{seesaw} \\
{\mathbbm m}_2 &\approx& {\mathbbm M}_{\rm N} \; ,
\label{seesaw-heavy}
\eeqa
where we have assumed that ${\mathbbm M}_{\rm N} \gg {\mathbbm m}_D$, {\it i.e.} the 
eigenvalues of ${\mathbbm M}_{\rm N}$ are much larger than the 
eigenvalues of ${\mathbbm m}_D$ and kept terms upto 
${\cal O}({\mathbbm m}_D/{\mathbbm M}_{\rm N})$. 
Thus Eqs.~(\ref{seesaw-heavy}) and (\ref{seesaw}) show respectively that 
eigenvalues of the matrix ${\mathbbm m}_2$ are large, while those of 
${\mathbbm m}_1$ are small and hence the eigenstates 
corresponding to these small eigenvalues should serve the purpose of the mass 
eigenstates of the light active neutrinos. Thus the presence of the heavy 
right-handed neutrinos will produce the light active neutrino masses and 
this mechanism of making one particle light at the expense of making another 
one heavy is called the seesaw mechanism. The seesaw obtained by 
adding these heavy right-handed singlet fermions to the SM is called the Type-I 
seesaw.

\begin{table}[t!] 
\begin{center} 
\begin{tabular}{|l|c|l|}
\hline
& & \\ & The effective vertex & $\quad \quad \quad \; \kappa$ \\
\hline
$\phantom{abc}$ & $\phantom{abc}$ & $\phantom{abc}$ \\
\centering Type-I 
& $\phantom{abc}$  & 
$\kappa = 2 \text{Y}_{\text N}^{\text T} 
{\mathbbm M}_{\text N}^{-1} \text{Y}_{\text N}$ \\ 
 $\phantom{abc}$ &\includegraphics[scale=0.7]{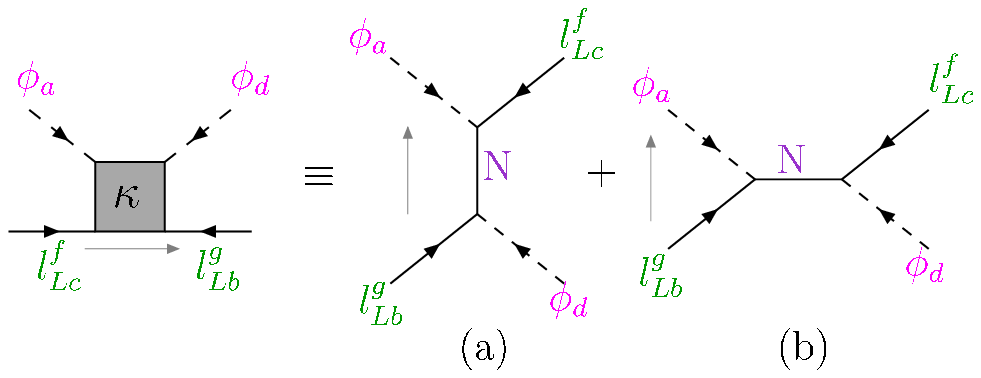} &$\phantom{abc}$ \\
\hline
$\phantom{abc}$ & $\phantom{abc}$ & $\phantom{abc}$ \\
\centering Type-II 
& $\phantom{abc}$  & 
$\kappa = -2 \frac{\text{Y}_{\Delta} \Lambda_6}
{{\mathbbm M}_{\Delta}^2}$ \\
 $\phantom{abc}$ &\includegraphics[scale=0.7]{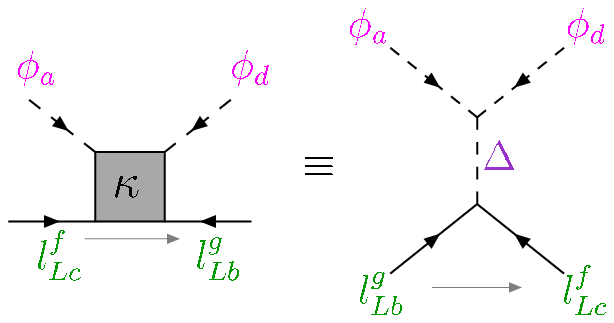} &$\phantom{abc}$ \\
\hline
$\phantom{abc}$ & $\phantom{abc}$ & $\phantom{abc}$ \\
\centering Type-III 
& $\phantom{abc}$  & 
$\kappa = 2 \text{Y}_{\Sigma}^{\text T} 
{\mathbbm M}_{\Sigma}^{-1} \text{Y}_{\Sigma}$ \\
 $\phantom{abc}$ &\includegraphics[scale=0.7]{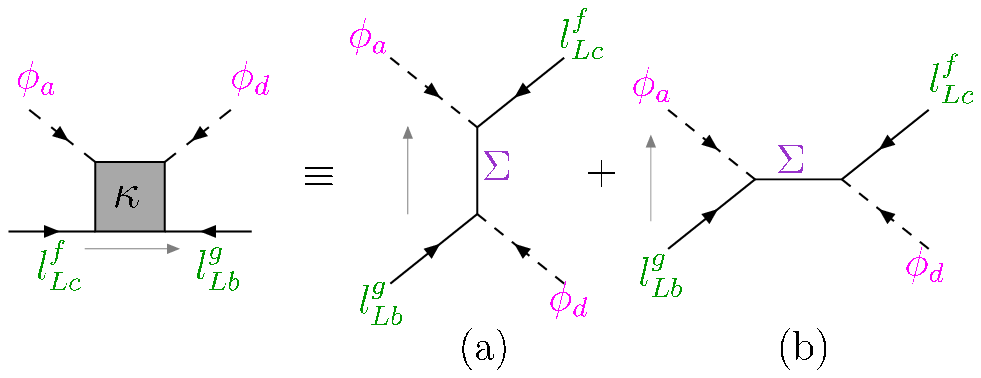} &$\phantom{abc}$ \\
\hline
\end{tabular}
\end{center}
\caption{Summary of the low energy effective couplings and the 
effective neutrino mass ${\mathbbm m}_\nu \equiv -\frac{v^2}{4} \kappa$ 
in the three seesaw scenarios. 
Here, $Y_{\rm N}$($Y_{\Sigma}$) are the Yukawa couplings for the heavy singlet(triplet) 
fermion present in Type-I(Type-III) seesaw and ${\mathbbm M}_{\rm N}$(${\mathbbm M}_{\Sigma}$) 
is the mass matrix ($N \equiv N_R + N_R^C$; $\Sigma \equiv \Sigma_R + \Sigma_R^C$). 
In Type-II seesaw, ${\mathbbm M}_{\Delta}$ is the mass of the heavy triplet Higgs, 
$Y_{\Delta}$ is its Yukawa coupling with the SM lepton doublet $l_L$, and 
$\Lambda_6$ is its coupling with the SM Higgs $\phi$.
\label{seesaw-summary}}
\end{table}

In the low energy limit we have an effective theory 
described by \cite{weinberg}
\beqa 
{\cal L}_\kappa & = & \kappa_{fg}  
\left( {\overline{l_L^C}}^f \sigma^i \varepsilon  \phi \right)   
\left( \phi^T \sigma^i \varepsilon l_L^g \right) + {\rm h.c.} ,   
\\ 
&=& - \kappa_{fg} \left( {\overline{l_L^C}}^f_c  
\phi_a  l_{Lb}^g \phi_d \right)  
\frac{1}{2} \left( \varepsilon_{ac} \varepsilon_{bd} +  
\varepsilon_{ab} \varepsilon_{cd} \right) + {\rm h.c.} \; \; , 
\label{L-kappa} 
\eeqa 
where $\kappa$ is a symmetric complex matrix with mass dimension  
$(-1)$ and $\varepsilon \equiv i \sigma^2$ is the completely anti-symmetric 
tensor in the SU(2)$_L$ space. Generation indices $f,g \in \{ 1,2,3\}$ are shown  
explicitly and $a, b, c, d \in \{1,2\}$ are the SU(2)$_L$ indices.  
In writing Eq.~(\ref{L-kappa}) we have used 
\begin{eqnarray} 
(\sigma^i)_{ab} (\sigma^i)_{cd} &=& 2 \delta_{ad} \delta_{bc} - \delta_{ab} \delta_{cd} \nn \\ 
\Rightarrow \quad  
(\sigma^i \varepsilon)_{ba} (\sigma^i \varepsilon)_{dc} &=& 
2 \varepsilon_{da} \varepsilon_{bc} - \varepsilon_{ba} \varepsilon_{dc} 
\label{su2-algebra} 
\end{eqnarray} 
and utilizing the $\phi_d \leftrightarrow \phi_a$ symmetry, we can write 
\begin{equation} 
2 \varepsilon_{da} \varepsilon_{bc} - \varepsilon_{ba} \varepsilon_{dc} 
= \frac{1}{2} \left( \varepsilon_{ab} \varepsilon_{dc} +  
\varepsilon_{db} \varepsilon_{ac} \right) \; . 
\label{symmetrize} 
\end{equation}
 
The relevant diagrams in the complete theory giving rise to  
the effective operators in the low energy limit are shown in the 
topmost row in the Table~\ref{seesaw-summary}. 
The ``shaded box'' on the left hand side of the equivalence in the 
middle column represents the effective low energy vertex $\kappa$, 
while ${\cal A}_{(a)}$ and ${\cal A}_{(b)}$ are the amplitudes of the  
diagrams labeled as $(a)$ and $(b)$ on the right hand side. 
The amplitudes are given by 
\beqa 
{\cal A}_{(a)} &=& i \mu^{\epsilon} \left( Y_{\rm N}^T{\mathbbm M}_{\rm N}^{-1}  
Y_{\rm N} \right)_{fg} \varepsilon_{ca} \varepsilon_{bd} P_L \; , \\ 
{\cal A}_{(b)} &=& i \mu^{\epsilon} \left( Y_{\rm N}^T {\mathbbm M}_{\rm N}^{-1}  
Y_{\rm N} \right)_{fg} \varepsilon_{cd} \varepsilon_{ba} P_L \; ,  
\eeqa 
with $\epsilon = 4 - D$ where $D$ is the dimensionality 
that we introduce in order to use dimensional regularization. 
Note that ${\cal A}_{(b)}$ is obtained from ${\cal A}_{(a)}$ just by  
$d \leftrightarrow a$ interchange. 
Using Eq.~(\ref{su2-algebra}) one  
finally gets 
\beqa 
{\cal A}_{(a)} + {\cal A}_{(b)} &=& -i \mu^{\epsilon}  
\left(Y_{\rm N}^T {\mathbbm M}_{\rm N}^{-1} Y_{\rm N} \right)_{fg} 
\left( \varepsilon_{ab}\varepsilon_{cd}  
+ \varepsilon_{ac}\varepsilon_{bd} \right) P_L \; . 
\label{kappa-rhs-I} 
\eeqa 
This is equal to the left hand side of the figure mentioned with the identification  
\beqa 
\kappa = 2 Y_{\rm N}^T {\mathbbm M}_{\rm N}^{-1} Y_{\rm N} \; , 
\label{kappaI} 
\eeqa 
as shown in the Table~\ref{seesaw-summary}.
From Eqs.~(\ref{L-kappa}) and (\ref{kappaI}), one gets  
the neutrino mass after spontaneous symmetry breaking to be  
\beq 
{\mathbbm m}_\nu = -\frac{v^2}{2}  Y_{\rm N}^T {\mathbbm M}_{\rm N}^{-1} Y_{\rm N} 
\eeq  
which is the Type-I seesaw relation.  
As the energy changes, the heavy singlets get decoupled one by one at their 
respective mass scales and start contributing to the light neutrino 
mass through the effective operator.


\subsubsection{Type-II seesaw}
\label{sec:typeII}

In the Type-II seesaw, we consider the SM extended by a charged 
Higgs triplet transforming in the adjoint representation of SU(2)$_L$
\begin{equation}
\Delta=\frac{\sigma^i\Delta^i}{\sqrt{2}}=\left(\begin{array}{cc}
    \Delta^+/\sqrt{2} & \Delta^{++}\\
    \Delta^0 & -\Delta^+/\sqrt{2}\\
  \end{array}\right)\; ,
\end{equation}
where $\Delta^{++} \equiv (\Delta^1 - i \Delta^2)/\sqrt{2}$, 
$\Delta^0 \equiv (\Delta^1 + i \Delta^2)/\sqrt{2}$ and 
$\sigma^i \equiv \{ \sigma^1, \sigma^2, \sigma^3 \}$ are the Pauli matrices. 
Following the notation of \cite{Schmidt-TypeII,Schmidt-thesis}, the Lagrangian is given by
\beq
{\cal L} = {\cal L}_\text{SM} + {\cal L}_\Delta \; ,
\eeq
where 
\beqa
{\cal{L}}_{\Delta} &=& {\cal{L}}_{\Delta,kin} + {\cal{L}}_{\Delta,\phi} + 
{\cal{L}}_{\Delta,Yukawa} \; .
\label{LDelta} 
\eeqa
Here
\beqa
{\cal{L}}_{\Delta,kin} &=& \Tr \left[\left( D_\mu  \Delta \right)^\dagger D^\mu\Delta\right] \; , 
\label{LDelta_kin} \\
{\cal{L}}_{\Delta,\phi} &=&- {\mathbbm M}_\Delta^2 \Tr \left(\Delta^\dagger \Delta\right) 
- \frac{\Lambda_1}{2}\left[\Tr \left( \Delta^\dagger \Delta\right) \right]^2 \nn \\ 
&&-  \frac{\Lambda_2}{2} \left[ 
\left[ \Tr \left( \Delta^\dagger \Delta \right) \right]^2 - 
\Tr \left( \Delta^\dagger \Delta \Delta^\dagger \Delta \right) \right] 
- \Lambda_4 \phi^\dagger \phi \Tr \left( \Delta^\dagger \Delta \right) \nn \\ 
&&- \Lambda_5 \phi^\dagger \left[ \Delta^\dagger,\Delta \right]\phi -
\left[ \frac{\Lambda_6}{\sqrt{2}} \phi^T i\sigma_2 \Delta^\dagger \phi + {\rm h.c.} \right] \; , 
\label{LDelta_phi} 
\eeqa
\beqa
{\cal{L}}_{\Delta,Yukawa} &=& 
- \frac{1}{\sqrt{2}}\left(Y_\Delta\right)_{fg}\ell_L^{Tf} {\rm C} (i\sigma_2) \Delta \ell_L^g + {\rm h.c.} \; ,
\label{LDelta_Yuk}
\eeqa
where $\mathrm{C}$ is the charge conjugation matrix with respect to the Lorentz group. 
The covariant derivative of the Higgs triplet is given by~\footnote{
We use GUT charge normalization: 
$\frac{3}{5} \left(g_1^\mathrm{GUT}\right)^2 = \left(g_1^\mathrm{SM}\right)^2$.}
\begin{equation}
D_\mu \Delta = \partial_\mu \Delta +i \sqrt{\frac{3}{5}} g_1 B_\mu \Delta + i g_2 \left[ W_\mu,\Delta\right] \; ,
\end{equation}
where $g_1$ and $g_2$ are the U(1)$_Y$ and SU(2)$_L$ gauge couplings respectively.
With the interactions shown in Eqs.~(\ref{LDelta})--(\ref{LDelta_Yuk}), 
after electroweak symmetry breaking the triplet Higgs 
$\Delta$ will get a vev given by $\langle \Delta_0 \rangle \sim \Lambda_6 v^2/ 2 \sqrt{2} M_\Delta^2$. 
This triplet Higgs vev will also contribute to the gauge boson masses and will alter 
the $\rho$-parameters from the SM prediction $\rho \approx 1$, at tree level and hence 
will get a strong constraint from the current precision data \cite{shafi-TypeII}.

Once the triplet Higgs $\Delta$ gets its vev after spontaneous symmetry breaking, 
the Lagrangian in Eq.~(\ref{LDelta_Yuk}) produces the neutrino mass term as
\beqa
{\cal{L}}_{\Delta,Yukawa} &=& \frac{1}{\sqrt{2}} Y_\Delta \nu_L^C 
\langle \Delta_0 \rangle \nu_L + {\rm h.c.} \; ,
\eeqa
and thus using the expression for $\langle \Delta_0 \rangle$, the neutrino mass is given as
\beqa
{\mathbbm m}_\nu &=& \frac{v^2}{2} \frac{\Lambda_6 Y_\Delta}{{\mathbbm M}_\Delta^2} \; .
\label{seesaw-II}
\eeqa

In case of Type-II seesaw, only one diagram in the complete high energy theory contributes 
to the effective low energy neutrino mass operator, as shown in Table~\ref{seesaw-summary}, 
and we have
\beqa
{\cal A} &=& i \frac{\Lambda_6}{{\mathbbm M}_\Delta^2} \left( Y_\Delta \right)_{fg} 
\left(   \varepsilon_{ac}\varepsilon_{bd} + \varepsilon_{ab} \varepsilon_{cd} \right) \; ,
\eeqa
and comparison with Eq.~(\ref{L-kappa}) gives
\beqa
\kappa = - \frac{2 \Lambda_6 Y_\Delta}{{\mathbbm M}_\Delta^2} \; .
\label{kappaII}
\eeqa
Hence finally one gets the neutrino mass to be  
\beq 
{\mathbbm m}_\nu = -\frac{v^2}{4} \kappa = \frac{v^2}{2} \frac{\Lambda_6 Y_\Delta}{{\mathbbm M}_\Delta^2} \; ,  
\eeq  
which is the same as the Type-II seesaw relation, as given in Eq.~(\ref{seesaw-II}).

Just like the right--handed neutrinos in case of Type-I seesaw, the Higgs 
triplets in Type-II seesaw will decouple step by step
at their respective mass scales and the effective theories have to be
matched against each other. The decoupling of the right--handed
neutrinos only contributes to the effective 5-dimensional neutrino mass operator, 
while the decoupling of the 
Higgs triplet also gives a contribution to the SM model Higgs self--coupling 
because there is a coupling between the SM Higgs doublet and the Higgs 
triplet given in Eq.~(\ref{LDelta_phi}). The matching condition 
for the Higgs self--coupling at the threshold is given as
\beqa
\lambda^{\rm EFT} &=& \lambda + 2\frac{|\Lambda_6|^2}{{\mathbbm M}_\Delta^2}\; .
\eeqa
%


\subsubsection{Type-III seesaw}
\label{sec:typeIII}

Type-III seesaw mechanism is mediated by heavy fermion triplets  
transforming in the adjoint representation of SU(2)$_L$ and 
has been considered earlier in \cite{Foot-Type-III,Ma:2002pf}. 
Very recently there has been a renewed interest in these type of models. 
The smallness of neutrino masses usually implies the mass of the 
heavy particle to be high $\sim 10^{11 - 15}$ GeV, as shown in 
Chapter~\ref{subsec:effective_L}. However, it is also possible that 
one or more of the triplets have masses near the TeV scale, 
making it possible to search for their signatures at the LHC 
\cite{goran-bajc-lhc,goran-bajc-triplet,Abada:2007ux,Franceschini:2008pz,delAguila:2008cj}. 
In such models, the Yukawa couplings need to be small to suppress the 
neutrino mass, if no fine tuning of the parameters is assumed.  
Lepton flavor violating decays in the context of Type-III seesaw  
models have also been considered in \cite{Abada:2008ea}.   
Recently it has also been suggested that the neutral member of the triplet  
can serve as the dark matter and can be instrumental in generating  
small neutrino mass radiatively \cite{ma-dm}. 

In the Type-III seesaw, there are right handed fermionic triplet  
$\Sigma_R$ added to the SM at the high scale which is singlet under 
U(1)$_Y$, while transform as a triplet in the adjoint representation 
of SU(2)$_L$. This triplet can be represented as 
\beq 
\Sigma_R = \left( \barr{cc} 
\Sigma_R^0/\sqrt{2} & \Sigma_R^+ \\ 
\Sigma_R^- & -\Sigma_R^0/\sqrt{2} 
\earr \right)  \equiv  \frac{\Sigma_R^i \sigma^i}{\sqrt{2}} \; , 
\label{Sigma} 
\eeq 
where $\Sigma_R^\pm = {(\Sigma_R^1 \mp i \Sigma_R^2)}{\sqrt{2}} $.
For the sake of simplicity of further calculations, we combine 
$\Sigma_R$ with its CP conjugate $\Sigma_R^C$ to construct 
\beq
\Sigma \equiv \Sigma_R + \Sigma_R^C \; .\\
\eeq
Clearly, $\Sigma$ also transforms in the adjoint representation  
of SU(2)$_L$. 
Note that though formally $\Sigma = \Sigma^C$,  the 
individual elements of $\Sigma$ are not all Majorana particles. 
While the diagonal elements of $\Sigma$ are indeed 
Majorana spinors which represent the neutral component of $\Sigma$, the 
off-diagonal elements are charged Dirac spinors.

Introduction of this triplet field will introduce new terms in the Lagrangian. 
The net Lagrangian is  
\beqa 
{\cal L} = {\cal L}_{SM} +  {\cal{L}}_{\Sigma} \; ,
\label{lagrangian-total}  
\eeqa 
where 
\beqa 
{\cal{L}}_{\Sigma} = {\cal{L}}_{\Sigma,kin} + {\cal{L}}_{\Sigma,mass} + 
{\cal{L}}_{\Sigma,Yukawa} \; . 
\label{L-Sigma}
\eeqa 
Here, 
\beqa 
{\cal{L}}_{\Sigma,kin} & = & \Tr[\overline{\Sigma} i \D  \Sigma] \; , 
\label{L-Sigma-kin}\\
{\cal{L}}_{\Sigma,mass} &  = & -\frac{1}{2} \Tr[\Sb {\mathbbm M}_\Sigma \Sigma] \; , 
\label{L-Sigma-mass}\\ 
{\cal{L}}_{\Sigma,Yukawa} & = & 
-\overline{l_L} \sqrt{2} Y_\Sigma^\dagger 
\Sigma \widetilde{\phi} - 
\phi^T \varepsilon ^T \Sb \sqrt{2} Y_\Sigma l_L \; . 
\label{L-Sigma-Yukawa} 
\eeqa 
Here we have not written the generation indices explicitly.  
${\mathbbm M}_\Sigma$ is the Majorana mass matrix of the heavy  
fermion triplets and $Y_\Sigma$ is the Yukawa coupling.  
Since the fermion triplet $\Sigma$ is in the adjoint  
representation of SU(2)$_L$, the covariant derivative  
of $\Sigma$ is defined as 
\beqa 
D_{\mu} \Sigma   = \partial_\mu \Sigma + i g_2 [W_\mu, \Sigma] \; , 
\eeqa 
where $g_2$ is the SU(2)$_L$ gauge coupling. Unlike $\Delta$, $\Sigma$ being 
a singlet under U(1)$_Y$ does not couple to $B_\mu$.

The new term ${\cal L}_{\Sigma}$ in the Lagrangian, as shown in Eq.~(\ref{L-Sigma}), 
can be expanded as 
\cite{Abada:2008ea}
\beqa
{\cal L}_{\Sigma} &=& 
\left( \overline{\Psi}i\del\Psi + \overline{\Sigma^0_R} i\del \Sigma^0_R + 
{\rm h.c.}\right) \nn \\ 
&&+ g_2 \left( W_\mu^+\overline{\Sigma^0_R}\gamma^\mu P_R \Psi 
+ W_\mu^+\overline{\Sigma^{0C}_R}\gamma^\mu P_L \Psi + {\rm h.c.} \right)
- g_2 W_\mu^3 \overline{\Psi} \gamma^\mu \Psi \nn \\
&&- \overline{\Psi} {\mathbbm M}_\Sigma \Psi 
- \left( \frac{1}{2} \overline{\Sigma^0_R} {\mathbbm M}_\Sigma \Sigma^{0C}_R + {\rm h.c.} \right)\nn \\
&& - \left( \phi^0 \overline{\Sigma^0_R} Y_\Sigma \nu_L + \sqrt{2} \phi^0 \overline{\Psi}Y_\Sigma l_L 
+ \phi^+ \overline{\Sigma^0_R} Y_\Sigma l_L - \sqrt{2} \phi^+ \overline{\nu_L^C}Y_\Sigma^T \Psi + {\rm h.c.} \right) \; .
\label{L-Sigma-expanded} \nn \\
\eeqa 
Here we have defined the four component Dirac spinor
\beq
\Psi \equiv \Sigma_R^{+C} + \Sigma_R^{-} \; , 
\eeq
for our convenience, while the neutral component of $\Sigma_R$ is 
still in the two component notation. 
In Eq.~(\ref{L-Sigma-expanded}), the first two lines come from ${\cal L}_{\Sigma,kin}$, the 
third line corresponds to the Majorana mass term in ${\cal L}_{\Sigma,mass}$ and the terms in 
the last line corresponds to the Yukawa coupling terms in ${\cal L}_{\Sigma,Yukawa}$, 
as given in Eqs.~(\ref{L-Sigma-kin})--(\ref{L-Sigma-Yukawa}). After 
the electroweak symmetry breaking, the mass matrix for the neutral 
fields become
\beqa
{\cal L} \ni 
-\frac{1}{2}\left( \barr{cc} \overline{\nu_L^C} & \overline{\Sigma^{0}_R} \earr \right) 
\left( \barr{cc} 0 & {\mathbbm m}_D \\ {\mathbbm m}_D^T & {\mathbbm M}_\Sigma \earr \right)
\left( \barr{c} \nu_L \\ \Sigma^{0C}_R \earr \right) + {\rm h.c.} \; ,
\label{neutral-mass}
\eeqa
where ${\mathbbm m}_D = (v/\sqrt{2}) Y_{\Sigma}^T$ is the Dirac mass matrix of the neutral fields. 
Thus the mass matrix in Eq.~(\ref{neutral-mass}) looks the same as that obtained in 
Eq.~(\ref{mass-matrix-nu}) and hence for large ${\mathbbm M}_\Sigma$, 
diagonalization of the mass matrix will 
produce light active neutrino states via seesaw mechanism, as 
obtained in Sec~\ref{sec:typeI}. The seesaw achieved here with the 
help of the neutral component of the fermionic triplet is known as the 
Type-III seesaw mechanism. 
Eq.~(\ref{neutral-mass}) also implies that there will be a mixing between 
the light and the heavy neutral states, however the mixing angle 
will be ${\cal O}({\mathbbm m}_D/{\mathbbm M}_\Sigma)$ and hence very small 
for large ${\mathbbm M}_\Sigma$.

Since the heavy fermion triplets added to the SM at the high scale 
have charged components also, 
they will modify the masses of the charged leptons belonging to the SM, 
in addition to the generation of the small active neutrino masses. With 
the addition of the triplet fields, the mass term of the charged 
lepton sector after electroweak symmetry breaking becomes
\beqa
{\cal L} \ni 
&-&\left( \barr{cc} \overline{l_R} & \overline{\Psi_R} \earr \right) 
\left( \barr{cc} {\mathbbm m}_L & 0 \\ \sqrt{2}{\mathbbm m}_D^T & {\mathbbm M}_\Sigma \earr \right)
\left( \barr{c} l_L \\ \Psi_L \earr \right)  \nn \\
&-& \left( \barr{cc} \overline{l_L} & \overline{\Psi_L} \earr \right) 
\left( \barr{cc} {\mathbbm m}_L & \sqrt{2}{\mathbbm m}_D^\ast \\ 0 & {\mathbbm M}_\Sigma \earr \right)
\left( \barr{c} l_R \\ \Psi_R \earr \right) \nn \\
= &-&\left( \barr{cc} \overline{l_R} & \overline{\Psi_R} \earr \right) 
{\cal M}_c \left( \barr{c} l_L \\ \Psi_L \earr \right) 
- \left( \barr{cc} \overline{l_L} & \overline{\Psi_L} \earr \right) 
{\cal M}_c^\dagger \left( \barr{c} l_R \\ \Psi_R \earr \right)\; , 
\label{charged-lepton-mass}
\eeqa
where ${\mathbbm m}_L$ is the Dirac mass matrix of the SM charged leptons and 
\beqa
{\cal M}_c \equiv \left( \barr{cc} {\mathbbm m}_L & 0 \\ 
\sqrt{2}{\mathbbm m}_D^T & {\mathbbm M}_\Sigma \earr \right)
\label{M-charged}
\eeqa
denotes the complete mass matrix for the charged leptons. Eq.~(\ref{M-charged}) 
shows that the inclusion of the charged fermions as components 
of the heavy triplets does not change the masses of the 
charged leptons of the SM upto  the order ${\cal O}(({\mathbbm m}_D, {\mathbbm m}_L)/{\mathbbm M}_\Sigma)$. 
However, there will be mixing between the states $l_L$-$\Psi_L$ and $l_R$-$\Psi_R$, 
but the mixing angle is small in the 
${\mathbbm m}_D,{\mathbbm m}_L \ll {\mathbbm M}_\Sigma$ limit. The 
correction to the charged lepton masses due to the charged components 
of the triplet fermion in ${\cal O}([({\mathbbm m}_D, {\mathbbm m}_L)/{\mathbbm M}_\Sigma]^2)$ 
can be calculated easily from \cite{Abada:2007ux}.

As can be seen from the Table~\ref{seesaw-summary}, the diagrams 
in the complete Type-III seesaw theory giving rise to the effective operators 
in the low energy limit are very similar 
to the case of Type-I seesaw. Here the amplitudes 
${\cal A}_{(a)}$ and ${\cal A}_{(b)}$ are given by 
\beqa 
{\cal A}_{(a)} &=& i \mu^{\epsilon} \left( Y_\Sigma^T{\mathbbm M}_\Sigma^{-1}  
Y_\Sigma \right)_{fg} \left[ (\varepsilon^T \sigma^i)_{ab}  
(\varepsilon^T \sigma^i)_{cd} \right] P_L \; , \\ 
{\cal A}_{(b)} &=& i \mu^{\epsilon} \left( Y_\Sigma^T {\mathbbm M}_\Sigma^{-1}  
Y_\Sigma \right)_{fg} \left[ (\varepsilon^T \sigma^i)_{db}  
(\varepsilon^T \sigma^i)_{ca} \right] P_L \; .  
\eeqa 
Using Eq.~(\ref{su2-algebra}) one finally gets 
\beqa 
{\cal A}_{(a)} + {\cal A}_{(b)} &=& -i \mu^{\epsilon}  
\left(Y_\Sigma^T {\mathbbm M}_\Sigma^{-1} Y_\Sigma \right)_{fg} 
\left( \varepsilon_{ab}\varepsilon_{cd}  
+ \varepsilon_{ac}\varepsilon_{bd} \right) P_L \; , 
\label{kappa-rhs-III} 
\eeqa 
which gives 
\beqa 
\kappa = 2 Y_\Sigma^T {\mathbbm M}_\Sigma^{-1} Y_\Sigma \; . 
\label{kappaIII} 
\eeqa 
From Eqs.~(\ref{L-kappa}) and ~(\ref{kappaIII}) one gets  
the neutrino mass after spontaneous symmetry breaking to be  
\beq 
{\mathbbm m}_\nu = -\frac{v^2}{2}  Y_\Sigma^T {\mathbbm M}_\Sigma^{-1} Y_\Sigma 
\eeq  
which is the Type-III seesaw relation.  
Here, $v$ denotes the vacuum expectation value of the  
Higgs  field. 


\subsubsection{Inverse seesaw}
\label{sec:inverse_seesaw}

Apart from the three types of seesaws described in 
Sec.~\ref{sec:typeI}--\ref{sec:typeIII}, there is another well-known 
scenario known as the Inverse Seesaw. 
In the Inverse Seesaw scenario \cite{Mohapatra:1986bd}, 
additional SM gauge singlets are introduced together with a small Majorana mass 
insertion through the additional right handed heavy singlets which explicitly 
breaks the lepton number. The minimal version of this Inverse Seesaw scenario 
requires the addition of two right-handed neutrinos $N_R^f$ and two 
left-handed SM gauge singlets $S_L^f$, where $f \in \{ 1,2\}$ is the 
generation index \cite{Inverse_Ohlsson-1,Inverse_Ohlsson-2}. We construct the field $N^f$ and $S^f$ as 
\beqa
N^f &=& N_R^f + N_R^{Cf} \; , \\
S^f &=& S_L^f + S_L^{Cf} \; , 
\eeqa
and then the Lagrangian for the theory will be given by
\beq
{\cal L} = {\cal L}_{\rm SM} + {\cal L}_{IS} \; ,
\eeq
where ${\cal L}_{IS}$ is the contribution from the fields added to the SM to 
produce the Inverse Seesaw and is given as
\beqa
{\cal L}_{IS} &=& {\cal L}_{IS,kin} +   {\cal L}_{IS,Yukawa} + {\cal L}_{IS,mass} \; , 
\eeqa
where
\beqa
{\cal L}_{IS,kin} &=& \frac{1}{2} \overline{N}^f \left( i\del \right)_{fg} N^g +  
\frac{1}{2} \overline{S^f} \left( i\del \right)_{fg} S^g \; , \\ 
{\cal L}_{IS,Yukawa} &=& -\overline{N}^f {\left(Y_{\rm N}\right)}_{fg} 
\widetilde{\phi}^\dagger l_L^g + {\rm h.c.} \; , \\
{\cal L}_{IS,mass} &=& - {\overline{S}}^f {\left({\mathbbm M}_R \right)}_{fg} N^g
- \frac{1}{2} {\overline{S}}^f \mu_{fg} S^{C g} + {\rm h.c.} \; ,
\eeqa
where the generation indices $g,f$ are written down explicitly. 
Here $\mu$ is a complex symmetric $2 \times 2$ matrix and $Y_{\rm N}$ 
and ${\mathbbm M}_{R}$ are arbitrary $3\times2$ and $2\times2$
matrices, respectively. Without loss of generality, one can always
redefine the extra singlet fields and work in a basis where $\mu$ is
real and diagonal. 
After the electroweak symmetry breaking, the Lagrangian giving rise to 
the $7 \times 7$ mass matrix for the neutral fields 
$\nu_L^f$, $N^f$ and $S^f$ can be expressed as
\beqa
{\cal L} \ni \left( \barr{ccc} \overline{\nu_L} & \overline{N_R^C} & 
\overline{S_L} \earr \right) 
\left( \barr{ccc} 0 & {\mathbbm m}_D & 0 \\
{\mathbbm m}_D^T & 0 & {\mathbbm M}_R^T \\
0 & {\mathbbm M}_R & \mu \earr \right)
\left( \barr{c} \nu_L^C \\ N_R \\ S_L^C \earr \right) + {\rm h.c.} \; ,
\eeqa
where the Dirac mass matrix is defined as ${\mathbbm m}_D = (v/\sqrt{2}) Y_{\rm N}^T$.
It should be noted that ${\mathbbm M}_R$ and $\mu$ being the mass terms of the 
SM singlet fields do not depend on the scale of the SU(2)$_L$ symmetry 
breaking. 

At the leading order in ${\mathbbm m}_D {\mathbbm M}_R^{-1}$, the active light 
neutrino mass matrix is given by
\beqa
{\mathbbm m}_\nu \approx {\mathbbm m}_D {\mathbbm M}_R^{-1} \mu 
{\left({\mathbbm M}_R^T\right)}^{-1} {\mathbbm m}_D^T \equiv F \mu F^T \; ,
\eeqa
where $F \equiv {\mathbbm m}_D {\mathbbm M}_R^{-1}$. In this case, 
for $\mu \sim 10^3$ eV, the light neutrino mass can be 
${\mathbbm m}_\nu \sim 0.01$ eV if $F \sim 0.3 \times 10^{-2}$. 
Thus for the Yukawa couplings $Y_{\rm N} \sim$ 0.1 -- 1, 
the heavy singlets can be in the mass range 
$10^4$--$10^5$ GeV and the seesaw scales can be lowered 
by orders of magnitudes compared to the Type-I seesaw.



\section{RG evolution of neutrino masses and mixing in effective theories}
\label{sec:RG_effective}

Now we consider the radiative corrections to the masses, mixing 
parameters and couplings 
in the effective low energy theory of neutrino masses.  
As can be understood from the discussions in 
Sec.~\ref{sec:mass-generation}, 
there is a unique dimension-5 operator given in Eq.~(\ref{dim-5 L}), 
that gives rise to the light neutrino masses after spontaneous 
symmetry breaking and hence is the 
same for all three types of seesaws.
Thus the RG evolution equations will depend solely on the 
underlying theory. Throughout this paper, we consider the SM as the 
low energy effective theory. However, to discuss the 
RG evolution in the effective theory, we will also 
consider the Minimal Supersymmetric Standard Model (MSSM) 
and discuss the effect of $\tan\beta$.

In order to evaluate the $\beta$-functions, one need to 
compute the renormalization constants for wavefunctions, 
couplings, etc. For this purpose any regularization scheme 
can be chosen from dimensional regularization, Pauli-Villars method, 
Ultra-Violet cutoff etc. and then the scheme for 
renormalization has to be fixed \cite{peskin}.
For the purpose of calculation of the renormalization constants, 
the gauge also has to be fixed. 
However, the final $\beta$-functions must be independent 
of the particular regularization as well as the 
renormalization scheme used for the calculations, 
and also of the gauge choice. 
$\beta$-functions can be determined from the renormalization 
constants using the functional differentiation method, as described in 
\cite{kersten-thesis,antusch}. 
The running equations for the Yukawa couplings, gauge couplings, 
Higgs self-coupling (in case of the SM only) and the effective neutrino 
mass operator in the SM and the MSSM are given in 
Table~\ref{effective-running}.
\begin{table}[t!] 
\begin{center} 
\small{
\begin{tabular}{|c|c|}
\hline
 & SM \\
\hline
$16 \pi^2 \beta_{\kappa}$
& $-\frac{3}{2}\left( Y_e^\dagger Y_e \right)^T \kappa -
\frac{3}{2} \kappa \left( Y_e^\dagger Y_e \right) + 
\left( 2 T + \lambda - 3 g_2^2 \right) \kappa$ \\
$16 \pi^2 \beta_{Y_e}$ 
& $Y_e \left( \frac{3}{2} Y_e^\dagger Y_e  
+ T - \frac{9}{4} g_1^2  - \frac{9}{4} g_2^2 \right)$ \\
$16 \pi^2 \beta_{Y_u}$
& $ Y_u \left( \frac{3}{2} Y_u^\dagger Y_u  
- \frac{3}{2} Y_d^\dagger Y_d + T  
-\frac{17}{20} g_1^2 - \frac{9}{4} g_2^2 - 8 g_3^2 \right)$ \\
$16 \pi^2 \beta_{Y_d}$ 
& $Y_d \left( \frac{3}{2} Y_d^\dagger Y_d  
- \frac{3}{2} Y_u^\dagger Y_u + T  
-\frac{1}{4} g_1^2 - \frac{9}{4} g_2^2 - 8 g_3^2 \right)$ \\
$16 \pi^2 \beta_\lambda$ & 
$6 \lambda^2 - 3 \lambda  
\left( \frac{3}{5} g_1^2 + 3 g_2^2 \right) 
+ 3 g_2^4 + \frac{3}{2} \left( \frac{3}{5} g_1^2 + g_2^2 \right)^2  
+ 4 \lambda T  - 8 T^\prime$ \\ 
$16 \pi^2 g_i$ 
& $b_i g_i^3$ ($b_1 = \frac{41}{10}$, $b_2 = -\frac{19}{6}$, $b_3 = -7$) \\
$T$ & $\Tr \left[ Y_e^\dagger Y_e + 3 Y_u^\dagger Y_u + 3 Y_d^\dagger Y_d \right]$ \\
$T^\prime$ & $\Tr [Y_e^\dagger Y_e Y_e^\dagger Y_e  
+ 3 Y_u^\dagger Y_u Y_u^\dagger Y_u  
+ 3 Y_d^\dagger Y_d Y_d^\dagger Y_d]$ \\
\hline
\hline
 & MSSM \\
\hline
$16 \pi^2 \beta_{\kappa}$
& $\left( Y_e^\dagger Y_e \right)^T \kappa + 
\kappa \left( Y_e^\dagger Y_e \right) + 
\left( 2 T_1 - \frac{6}{5} g_1^2 - 6 g_2^2 \right) \kappa$ \\ 
$16 \pi^2 \beta_{Y_e}$ 
& $Y_e \left( 3Y_e^\dagger Y_e  
+ T_2 - \frac{9}{5} g_1^2  - 3 g_2^2 \right)$ \\ 
$16 \pi^2 \beta_{Y_u}$
& $ Y_u \left( 3 Y_u^\dagger Y_u  + Y_d^\dagger Y_d + T_1  
-\frac{13}{15} g_1^2 - 3 g_2^2 - \frac{16}{3} g_3^2 \right)$ \\
$16 \pi^2 \beta_{Y_d}$ 
& $Y_d \left( 3 Y_d^\dagger Y_d  + Y_u^\dagger Y_u + T_2  
-\frac{7}{15} g_1^2 - 3 g_2^2 - \frac{16}{3} g_3^2 \right)$ \\  
$16 \pi^2 g_i$ 
& $b_i g_i^3$  ($b_1 = \frac{33}{5}$, $b_2 = 1$, $b_3 = -3$) \\
$T_1$ & $\Tr \left[ 3 Y_u^\dagger Y_u \right]$ \\
$T_2$ & $\Tr \left[ Y_e^\dagger Y_e + 3 Y_d^\dagger Y_d \right]$ \\
\hline
\end{tabular}
}
\end{center}
\caption{Evolution equations for the Yukawa couplings $Y_e$, $Y_u$, $Y_d$, 
gauge couplings $g_1$, $g_2$, $g_3$, Higgs self-coupling $\lambda$ (SM only) and 
the effective neutrino mass operator $\kappa$ in the SM and the MSSM 
\cite{kersten-thesis,antusch-HDM-MSSM,chankowski-pokorski,antusch-threshold}. 
Here we have used the GUT charge renormalization and hence 
$g_1 \equiv g_1^{\rm Unified} = \sqrt{5/3} \; g_1^{\rm SM}$. 
This convention has been followed through out this review. 
Here $\beta_X \equiv \mu (dX / d \mu)$. 
\label{effective-running} }
\end{table}
Finally, after electroweak symmetry breaking, the light neutrino mass matrix is given by 
\beq
{\mathbbm m}_\nu = -\frac{v^2}{4} \kappa   \; ,
\eeq
and thus will have the same evolution as $\kappa$. To simplify 
later discussions, the evolution equation for ${\mathbbm m}_\nu$ can be 
expressed as
\beqa
16 \pi^2 \beta_{{\mathbbm m}_\nu} &=& 
P^T {\mathbbm m}_\nu + {\mathbbm m}_\nu P + \alpha_\nu {\mathbbm m}_\nu 
\label{beta-mnu}
\eeqa
where $\beta_X \equiv d X/d \ln (\mu/ {\rm GeV})$ and
\beqa
P &=& C_e Y_e^\dagger Y_e\; . 
\label{P-nu-eff}
\eeqa
Here the values of $C_e$ and $\alpha_\nu$ depend on the underlying theory, 
and can be read off from Table~\ref{effective-running} when the theory is the
SM or the MSSM.
Without any loss of 
generality we can always choose the charged lepton Yukawa matrix $Y_e$ 
as well as the quark Yukawa matrices $Y_u$ and $Y_d$ to be diagonal at the 
high scale. Then from the RG equations in Table~\ref{effective-running} 
we get that they will remain diagonal at all energy scales, and so will $P$. 
Thus the evolution of the components of $\kappa$, and hence of 
${\mathbbm m}_\nu$, will be proportional to themselves.


\subsection{Evolution equations for neutrino parameters from matrix equations}
\label{subsec:effective_parameter_RG}

At any energy scale $\mu$, the neutrino mass matrix  
${\mathbbm m}_\nu$ and the charged lepton Yukawa $Y_e^\dagger Y_e$ 
can be diagonalized by unitary transformations via \cite{antusch-threshold}  
\beqa
U_\nu(\mu)^T {\mathbbm m}_\nu(\mu) U_\nu(\mu) &=& {\rm Diag}(m_1(\mu), m_2(\mu), m_3(\mu)) \; , 
\label{mnu-diag} \\
U_e(\mu)^\dagger Y_e^\dagger Y_e(\mu) U_e(\mu) &=& {\rm Diag}(y_e^2(\mu), y_\mu^2(\mu), y_\tau^2(\mu)) \; ,
\label{Ue-def}
\eeqa  
where $U_\nu$ and $U_e$ are unitary matrices and the neutrino mixing matrix 
will then be given by
\beq
U_{\rm PMNS}(\mu) = U_e^\dagger(\mu) U_\nu(\mu) \; .
\label{Upmns-Ue-Unu}
\eeq
Since in the effective theory $Y_e^\dagger Y_e$ remains diagonal at all energies, as 
discussed in the previous section, $U_e(\mu) = {\mathbbm 1}$ in Eq.~(\ref{Ue-def}) 
and from Eq.~(\ref{Upmns-Ue-Unu}) one has $U_{\rm PMNS}(\mu) = U_\nu(\mu)$. 
Thus the RG evolution of 
$U_{\rm PMNS}(\mu)$ will be governed by the running of ${\mathbbm m}_\nu$ only.
The evolution of the mixing matrix $U_\nu$ will be given by \cite{antusch-CP} 
\beq
\frac{d U_{\nu}}{dt} = U_{\nu} T \; ,
\label{RG-U}
\eeq
where $t \equiv \ln (\mu/{\rm GeV})/16 \pi^2$ and $T$ is an anti-Hermitian 
matrix defined as \cite{antusch-CP}
\beqa
16 \pi^2 \; {\rm Re} T_{ij} &=& 
\left\{ \barr{ll}
0 & (i = j) \; , \\
-\frac{m_i + m_j}{m_i - m_j} \; {\rm Re} P^\prime_{ij} & (i \ne j) \; , 
\earr \right.
\label{def-ReT}\\
16 \pi^2 \; {\rm Im} T_{ij} &=& -\frac{m_i - m_j}{m_i + m_j} \; {\rm Im} P^\prime_{ij} \; ,
\label{def-ImT} 
\eeqa
where $P^\prime = U_\nu^\dagger P U_\nu$, with $P$ defined by 
Eqs.~(\ref{beta-mnu})-(\ref{P-nu-eff}). In order to obtain the RG evolution 
equations for the mixing angles and the phases, one has to solve the 
system of nine coupled equations in the parameters 
$\xi_k = \{ \theta_{12}, \theta_{23}, \theta_{13}, \delta, \phi_1, \phi_2, \delta_e, 
\delta_\mu, \delta_\tau \}$, obtained from Eq.~(\ref{RG-U}) 
using the definition of $T$ as given in Eqs.~(\ref{def-ReT})-(\ref{def-ImT}) and 
then using the parametrization of $U_{\rm PMNS}$ as given in 
Eqs.~(\ref{Upmns})-(\ref{U}).


\subsubsection{RG evolution of the mixing angles and phases}
\label{sec:RG-mixingparam_effective}

The RG evolution equations for the mixing 
angles can be written in general as 
\cite{kersten-thesis,antusch-threshold,antusch-CP}
\beq
\dot{X} = \frac{D_X}{\theta_{13}} + A_X + {\cal O}(\theta_{13}) \; ,
\eeq
where $X \in \{  \theta_{12}, \theta_{23}, \theta_{13}, \delta, \phi_1, \phi_2 \}$. 
The differentiation 
is performed w.r.t. $t \equiv \ln (\mu/{\rm GeV})/16 \pi^2$. It can be seen that 
the quantities $D_X = 0$ for all $X$ except $D_{13}$. 
Evolution of the mixing angles are given by
\beqa
A_{12} &=& - \frac{C y_\tau^2}{2} \sin{2\theta_{12}} s_{23}^2 
\frac{\left\arrowvert m_1 e^{2 i \phi_1} + m_2 e^{2 i \phi_2} \right\arrowvert^2}{\dms} \; ,
\label{A12} \\
A_{23} &=& - \frac{C y_\tau^2}{2} \sin{2\theta_{23}} 
\left[ c_{12}^2 \frac{\left\arrowvert m_2 e^{2 i \phi_2} + m_3 \right\arrowvert^2}{\dma} 
+ s_{12}^2 \frac{\left\arrowvert m_1 e^{2 i \phi_1} + m_3 \right\arrowvert^2}{\dma(1+\zeta)} 
\right] \; , \quad \quad
\label{A23} \\
A_{13} &=& \frac{C y_\tau^2}{2} \sin{2\theta_{12}} \sin{2\theta_{23}}
\frac{m_3}{\dma(1+\zeta)} \times \nn \\
&&\left[ m_1 \cos{(2 \phi_1 - \delta)} - 
(1+\zeta) m_2 \cos{(2 \phi_2 - \delta)} - \zeta m_3 \right] \; ,
\label{A13}
\eeqa
where $\zeta = \dms/\dma$ and $c_{ij} = \cos{\theta_{ij}}$, 
$s_{ij} = \sin{\theta_{ij}}$. As can be seen from Table~\ref{effective-running}, 
$C = C_e = -3/2$ for the SM and $C = C_e = 1$ for the MSSM. Here $y_e^2$ and $y_\mu^2$ 
are neglected compared to $y_\tau^2$. 
The quantities governing the evolution of the Dirac CP 
phase $\delta$ are given as
\beqa
A_\delta &=& 2 C y_\tau^2 \Bigl\{
\frac{m_1 m_2}{\dms} s_{23}^2 \sin{(2\phi_1 - 2 \phi_2)} \Bigr. \nn \\
&& \Bigl. + \frac{m_3}{\dma(1+\zeta)} \left[ 
c_{23}^2 \left( m_1 c_{12}^2 \sin{(2\delta - 2\phi_1)} 
+ m_2 (1+\zeta) s_{12}^2 \sin{(2\delta - 2\phi_2)} \right) \right. \Bigr. \nn\\
&& \Bigl. \left. + \cos{2 \theta_{23}}  
\left( m_1 s_{12}^2 \sin{2 \phi_1} 
+ m_2 (1+\zeta) c_{12}^2 \sin{2 \phi_2} \right) \right ] \Bigr\} \; , 
\label{Adelta}\\
%
D_\delta &=& \frac{C y_\tau^2}{2} \sin{2\theta_{12}} \sin{2\theta_{23}}
\frac{m_3}{\dma(1+\zeta)} \times \nn \\
&&\left[ m_1 \sin{(2 \phi_1 - \delta)} - 
(1+\zeta) m_2 \sin{(2 \phi_2 - \delta)} + \zeta m_3 \sin\delta \right] \; ,
\label{Ddelta} 
\eeqa
while those for the Majorana phases $\phi_1$, $\phi_2$ are 
\beqa
A_{\phi_1} &=& 2 C y_\tau^2 \left\{ m_3 \cos{2 \theta_{23}} 
\frac{m_1 s_{12}^2 \sin{2\phi_1} +(1+\zeta) m_2 c_{12}^2 \sin{2\phi_2}}
{\dma(1+\zeta)} \right. \nn \\ 
&& \left. + \frac{m_1 m_2 c_{12}^2 s_{23}^2 \sin{(2\phi_1 - 2\phi_2)} }{\dms}\right\} \; , 
\label{Aphi1} 
\eeqa
\beqa
A_{\phi_2} &=& 2 C y_\tau^2 \left\{ m_3 \cos{2 \theta_{23}} 
\frac{m_1 s_{12}^2 \sin{2\phi_1} + (1+\zeta) m_2 c_{12}^2 \sin{2\phi_2}}
{\dma(1+\zeta)} \right. \nn \\ 
&& \left. + \frac{m_1 m_2 s_{12}^2 s_{23}^2 \sin{(2\phi_1 - 2\phi_2)}}{\dms}\right\} \; . 
\label{Aphi2} 
\eeqa


\subsubsection{RG evolution of the light neutrino masses}
\label{sec:RG-mass_effective}

Using the definition of $U_\nu$ in Eq.~(\ref{mnu-diag}) and $T$ in Eq.~(\ref{RG-U}), 
the RG evolution of the light neutrino masses is obtained from Eq.~(\ref{beta-mnu}) 
to be
\beq
\dot{m}_i = \left( {\rm Re}\alpha_\nu + 2 {\rm Re} P^\prime_{ii} \right) m_i \; ,
\eeq
where no summation over the repeated index `$i$' is to be taken. The 
evolution of the individual masses in terms of the mixing parameters 
becomes
\beqa
\dot{m}_1 &=& \left[ \alpha_\nu + C y_\tau^2 \left( 2 s_{12}^2 
s_{23}^2 + G_1 \right) \right] m_1 \; , 
\label{m1-dot}\\
\dot{m}_2 &=& \left[ \alpha_\nu + C y_\tau^2 \left( 2 c_{12}^2 
s_{23}^2 + G_2 \right) \right] m_2 \; , 
\label{m2-dot}\\
\dot{m}_3 &=& \left[ \alpha_\nu + 2 C y_\tau^2 c_{13}^2 
c_{23}^2 \right] m_3 \; ,
\label{m3-dot}
\eeqa
where
\beqa
G_1 &=& -s_{13} \sin2\theta_{12} \sin2\theta_{23} \cos\delta
+ 2 s_{13}^2 c_{12}^2 c_{23}^2 \; , \label{G1} \\
G_2 &=& s_{13} \sin2\theta_{12} \sin2\theta_{23} \cos\delta
+ 2 s_{13}^2 s_{12}^2 c_{23}^2 \; . \label{G2}
\eeqa
From the Eqs.~(\ref{m1-dot})--(\ref{m3-dot}) it can be seen that the 
evolution of a particular mass eigenvalue is proportional to itself 
upto ${\cal O}(\theta_{13}^0)$ and thus if some $m_i$ is zero to start with 
along with $\theta_{13} = 0$, it will remain so. However, it is 
the characteristic of the 1-loop RG evolution only, and breaks down 
when the 2-loop contributions are taken into account \cite{ishidori}.


To study the RG evolution of the neutrino masses and the mixing parameters 
in the effective theory, we consider $\mu_0$ to be the high energy scale below 
which the effective theory gives the correct description of the light 
neutrino masses, which we also take to be the mass of the lightest heavy
particle responsible for the seesaw mechanism. Then at any energy scale $\mu$, 
the value of the mixing angles can be expressed as
\beqa
\theta_{ij} &=& \theta_{ij}^0 + \int_{t_0}^t A_{ij}(t^\prime) dt^\prime + {\cal O}(\theta_{13})
\label{thij-int} \\
&\approx&  \theta_{ij}^0 +  k_{ij} \Delta_\tau + 
{\cal O}(\Delta_\tau \theta_{13},\Delta_\tau^2 ) \; ,
\label{thij-approx}
\eeqa
where $t_0 \equiv \ln (\mu_0/{\rm GeV})/16 \pi^2$ and $\theta_{ij}^0$ 
is the value of the angle at the high energy $\mu_0$. In Eq.~(\ref{thij-approx}), 
$\Delta_\tau$ is defined as
\beq
\Delta_\tau^{{\rm SM}} \equiv-\frac{1}{32 \pi^2} 
\left(\frac{g_2 m_\tau}{M_W}\right)^2
\ln{\left( \frac{\mu_0}{\mu} \right)}
\label{Delta_tau-SM}
\eeq
in the SM, where $g_2$ is the SU(2)$_L$ gauge coupling, whereas
$m_\tau$ and $M_W$ are the $\tau$ lepton and W boson 
masses respectively. In the MSSM,
\beq
\Delta_\tau^{{\rm MSSM}} \equiv-\frac{1}{32 \pi^2} 
\left(\frac{g_2 m_\tau}{M_W}\right)^2 (1+\tan^2{\beta}) 
\ln{\left( \frac{\mu_0}{\mu} \right)} \; .
\label{Delta_tau-MSSM}
\eeq
Numerically, one has $\Delta_\tau^{\rm SM} \approx -1.4 \times 10^{-5}$ when 
$\mu_0 = 10^{12}$ GeV and $\mu = 10^{2}$ GeV. For MSSM, 
$\Delta_\tau^{\rm MSSM} \approx -1.3 \times 10^{-5} (1 + \tan^2 \beta)$,
where $\mu = 10^{3}$ GeV and $\tan \beta$ can take values upto $\sim 50$. Hence 
in both the cases one can treat these quantities as small parameters. 
The quantities $k_{ij}$ can then be written from Eqs.~(\ref{A12})-(\ref{A13}) as
\beqa
k_{12} &=& - \frac{C}{2} \sin{2\theta_{12}} s_{23}^2 
\frac{\left\arrowvert m_1 e^{2 i \phi_1} + m_2 e^{2 i \phi_2} \right\arrowvert^2}{\dms} \; ,
\label{k12} \\
k_{23} &=& - \frac{C}{2} \sin{2\theta_{23}} 
\left[ c_{12}^2 \frac{\left\arrowvert m_2 e^{2 i \phi_2} + m_3 \right\arrowvert^2}{\dma} 
+ s_{12}^2 \frac{\left\arrowvert m_1 e^{2 i \phi_1} + m_3 \right\arrowvert^2}{\dma(1+\zeta)} 
\right] \; , \quad \quad
\label{k23} \\
k_{13} &=& \frac{C}{2} \sin{2\theta_{12}} \sin{2\theta_{23}}
\frac{m_3}{\dma(1+\zeta)} \times \nn \\
&&\left[ m_1 \cos{(2 \phi_1 - \delta)} - 
(1+\zeta) m_2 \cos{(2 \phi_2 - \delta)} - \zeta m_3 \right] \; .
\label{k13}
\eeqa
The same results are obtained in \cite{Dighe:2006zk,Dighe:2006sr,Dighe:2007ksa}
following a slightly different approach given in 
\cite{Ellis:1999my,Chankowski:1999xc,chankowski-pokorski}. Similar integrated 
evolution equations can be written for the Majorana phases as
\beqa
\phi_{i} &=& \phi_{i}^0 + k_{\phi_i} \Delta_\tau 
+ {\cal O}(\Delta_\tau \theta_{13}, \Delta_\tau^2) \; ,
\label{phi-int}
\eeqa
where $\phi_{i}^0$ is the value of $\phi_{i}$ at $\mu_0$ and 
$k_{\phi_i}$ can be read off directly from 
Eqs.~(\ref{Aphi1})-(\ref{Aphi2}). However, the running 
of the Dirac CP phase $\delta$ has to be considered carefully, 
since $D_\delta$ is non-zero and $\theta_{13}$ 
is allowed to take small values including zero. This issue will be discussed in 
detail in Sec~\ref{sec:subtlety}.

 
From $k_{12}$ in Eq.~(\ref{k12}) it can be seen that the solar mixing angle 
$\theta_{12}$ generically has the strongest RG effects
among the mixing angles. The reason for this is the smallness of the $\dms$ 
associated with it, in particular compared to $\dma$, which leads
to an enhanced running for quasi-degenerate neutrinos and for the case of 
an inverted mass hierarchy. The running is maximum for $|\phi_1 - \phi_2| = 0$, 
and minimum for $|\phi_1 - \phi_2| = \pi/2$. As it is clear from Eq.~(\ref{k12}), 
the direction of the running 
depends solely on $C \Delta_\tau$. Hence for evolution from a high to a 
low energy scale, $\theta_{12}$ always increases in the MSSM, and 
decreases in the SM. From Eq.~(\ref{k23}) it is evident that the 
direction of the $\theta_{23}$ evolution depends on $C \Delta_\tau$ as 
well as on the hierarchy. 
However, the running of $\theta_{13}$ depends on specific combinations 
of the CP phases, as shown in Eq.~(\ref{k13}). If the symmetry 
$\theta_{13} = 0$ is implemented at the high scale $\mu_0$ \cite{jcppaper}, 
which is the case for many neutrino mass models,  
the maximum $\theta_{13}$ value can be achieved with the choice 
\beq
2\phi_1 - \delta_0 = 0 \; , \quad 
|2\phi_2 - \delta_0| = \pi \; ,
\eeq
$\delta_0$ being the Dirac CP phase at $\mu_0$ and finally 
\beqa
\theta_{13}^{\rm max} &\leq & \frac{|C|  \Delta_\tau}{2} \sin{2 \theta_{12}} 
\sin{2 \theta_{23}} \frac{m_3}{|\dmsq_{31}|} 
 \bigl[ m_1+ (1+\zeta) m_2 + |\zeta| m_3  \bigr] 
\label{th13-max}\; .
\eeqa

To consider the running of the Majorana phases, one gets combining 
Eqs.~(\ref{Aphi1})-(\ref{Aphi2}) and Eq.~(\ref{phi-int})
\beqa
k_{\phi_1} - k_{\phi_2} &=& 2 C \cos{2 \theta_{12}} s_{23}^2 \frac{m_1 m_2}{\dms} 
\sin{(2\phi_1 - 2\phi_2)}\; , 
\label{phi-diff-run}
\eeqa
which shows that if $(\phi_1 - \phi_2) = 0$ at some scale, it will 
remain so at all energy scales, upto ${\cal O}(\theta_{13}^0)$. 
Moreover, if $(\phi_1 - \phi_2)$ is small at some scale so 
that we can write $\sin(2\phi_1 - 2\phi_2) \approx 2(\phi_1 - \phi_2)$, 
the running of $(\phi_1 - \phi_2)$ is proportional to itself.  
However, the ${\cal O}(\theta_{13})$ term may become important for 
large $\tan\beta$ values in case of the MSSM and then it will be possible 
to generate $(\phi_1 - \phi_2)$ radiatively. 

The running of the mass eigenvalues is significant even in the SM or 
for strongly hierarchical neutrino masses due to the factor $\alpha_\nu$ 
in the RG evolution equations given in Eqs.~(\ref{m1-dot})-(\ref{m3-dot}). 
As can be seen explicitly, the evolutions are not directly dependent 
on the Majorana phases, and the dependence on the Dirac CP phase $\delta$ 
is proportional to $\sin \theta_{13}$. Moreover, apart from the 
MSSM with very large $\tan \beta$ or at very high energy values, 
the running of the mass eigenvalues is 
solely controlled by the term proportional to $\alpha_\nu$, and there 
will be very small dependence on the mixing parameters. Thus in such cases, 
the running is given by a common scaling of the mass eigenvalues 
\cite{antusch-CP,chankowski-pokorski} and can be given by
\beqa
m_i \approx m_i^0 \; \exp\left[\int_{t_0}^t \alpha_\nu(t^\prime) dt^\prime \right] \; .
\label{int-mi}
\eeqa

Some generic features of the RG evolution of the light neutrino masses 
and the mixing parameters in the effective theory have been studied 
extensively in literature \cite{antusch-CP,chankowski-pokorski,babu-pantaleone,antusch, 
antusch-HDM-MSSM,Fukuyama:2002ch}.
These effects can have interesting consequences such as 
the generation of large mixing angles 
\cite{tanimoto,haba-LA,balaji-dighe, Balaji:2000ma,mpr,Agarwalla:2006dj},  
small mass splittings for degenerate neutrinos 
\cite{vissani-deg,branco-deg,Casas:1999tp, casas-deg,haba-deg,adhikari-deg, 
Joshipura:2002xa,Joshipura:2003fy,xing-deg,petcov-majorana}, or 
radiative generation of $\theta_{13}$ starting from a zero value 
at the high scale 
\cite{jcppaper,Joshipura:2002kj,Joshipura:2002gr,Mei:2004rn}. 
Some specific features of the RG evolution, like the stability of 
mixing angles and masses 
\cite{Ellis:1999my,lola,haba-stability,ma-stability,Haba:2000rf}, 
possible occurrence of fixed points 
\cite{chankowski-fixed,Pantaleone-fixed,xing-fixed} have also been studied.
RG induced deviations from various high scale  symmetries like 
tri-bimaximal mixing scenario 
\cite{Dighe:2006sr,Dighe:2007ksa,Plentinger:2005kx} or 
quark-lepton complementarity 
\cite{Dighe:2006zk,Dighe:2007ksa,Hirsch:2006je,Schmidt:2006rb,TBM-sray}  
and correlations with low scale observables have also been explored in detail.


\subsubsection{A subtlety at $\theta_{13} = 0$}
\label{sec:subtlety}

As mentioned already, Eq.~(\ref{Adelta}) clearly suggests that $A_\delta$ 
and hence $\dot{\delta}$ diverges for $\theta_{13} \to 0$.  
This problem is overcome by requiring that $D_\delta = 0$ at 
$\theta_{13} = 0$, which gives the 
following condition on $\delta$ at $\theta_{13} = 0$ \cite{antusch-CP}:
\beqa
\cot{\delta} = 
\frac{m_1 \cos{2\phi_1} - (1+\zeta) m_2 \cos{2 \phi_2} - \zeta m_3}
{m_1 \sin{2\phi_1} - (1+\zeta) m_2 \sin{2 \phi_2}} \; .
\label{cot-delta}
\eeqa
The above prescription works for the calculation of evolution
when one starts with vanishing $\theta_{13}$. However
on the face of it, it seems to imply that 
the CP phase $\delta$, which does not have any physical meaning 
at the point $\theta_{13} = 0$, should attain a particular value depending 
on the masses and Majorana phases, as given in Eq.~(\ref{cot-delta}). 
Moreover, getting the required value of $\delta$ precisely when 
$\theta_{13}=0$ would seem to need fine tuning when one starts 
from some non-zero $\theta_{13}$, unless this value of $\delta$ 
is a natural limit of the RG evolution when $\theta_{13} \to 0$.
The problem also propagates to the evolution of $\theta_{13}$,
since $A_{13}$ in Eq.~(\ref{A13}) depends in turn on $\delta$.
The evolution of all the other parameters, viz. $m_i$,
$\theta_{12}, \theta_{23}$ and $\phi_i$ is 
independent of $\delta$ upto ${\cal O}(\theta_{13}^0)$, 
and hence will have continuous, non-singular 
evolution even at $\theta_{13} = 0$.
This apparent singularity in $\delta$ has been explored in \cite{jcppaper} 
by analyzing the evolution of the complex quantity ${\cal U}_{e3}$, 
which stays continuous throughout the RG evolution 
and shows that a fine tuning is indeed required, but that is to 
ensure that $\theta_{13}$ exactly vanishes.
However, if the parameters happen to be tuned such that 
$\theta_{13}$ vanishes exactly, then the limiting value
of $\delta$ as $\theta_{13} \to 0$ is always the one given
by the prescription mentioned in Eq.~(\ref{cot-delta}). 

Even after understanding the origin of the apparent singularity 
in the evolution of $\delta$ and hence of $\theta_{13}$, a necessity 
still remains to have a clear evolution of parameters that
reflect the continuous nature of the evolution of elements
of the neutrino mixing matrix $U_{\rm PMNS}$. This can be achieved 
by choosing the basis as ${\cal P}_J = \{m_i, \theta_{12}, \theta_{23}, \theta_{13}^2, \phi_i, 
J_{\rm CP}, J'_{\rm CP} \}$ where the quantities $J_{\rm CP}$,$J^\prime_{\rm CP}$ 
are defined as
\beqa
J_{\rm CP} &=& \frac{1}{2} s_{12} c_{12} s_{23} c_{23} s_{13} c_{13}^2 \sin\delta \; , 
\label{Jcp} \\
J'_{\rm CP} &=& \frac{1}{2} s_{12} c_{12} s_{23} c_{23} s_{13} c_{13}^2 \cos\delta \; ,
\label{Jcp-prime}
\eeqa
instead of the conventional basis ${\cal P}_\delta \equiv 
\{m_i, \theta_{12}, \theta_{23}, \theta_{13}, \phi_i, \delta \}$. 
From Eqs.~(\ref{Jcp}) and (\ref{Jcp-prime}) it is seen that 
$J_{\rm CP}, J_{\rm CP}^\prime \to 0$ as $\theta_{13} \to 0$ and thus are 
well-defined.
The RG evolution equations for $J_{\rm CP}$ and $J'_{\rm CP}$ are given as
\beqa
\dot{J}_{\rm CP} &=& A_J + {\cal O}(\theta_{13}) \; , \\
\dot{J'}_{\rm CP} &=& A_J' + {\cal O}(\theta_{13}) \; ,
\eeqa
with
\beqa
A_J &=& C y_\tau^2 s_{12}^2 c_{12}^2 s_{23}^2 c_{23}^2 
\frac{m_3\Bigl[ m_1 \sin 2\phi_1 - (1+\zeta) m_2 \sin2\phi_2 \Bigr]}{\dma(1+\zeta)} 
\; , \label{AJ} \\
A_J^\prime &=& C y_\tau^2 s_{12}^2 c_{12}^2 s_{23}^2 c_{23}^2 
\frac{m_3\Bigl[ m_1 \cos 2\phi_1 - (1+\zeta) m_2 \cos2\phi_2 - \zeta m_3\Bigr]}{\dma (1+\zeta)} 
\; . \quad
\label{AJprime}
\eeqa 

In the new basis ${\cal P}_J$, the RG evolution for $\theta_{13}^2$ is considered 
instead of $\theta_{13}$, as is traditionally done. 
This quantity turns out to have a nonsingular behavior at $\theta_{13} = 0$.
Moreover, since $\theta_{13} \geq 0$ by convention, the complete 
information about $\theta_{13}$ lies within $\theta_{13}^2$. Also, 
the possible ``sign problem''\footnote{Usually 
the convention used in defining the elements of $U_{\rm PMNS}$ is to take 
the angles $\theta_{ij}$ to lie in the first quadrant. 
${\cal U}_{e3}$ can then take 
both positive or negative values depending on the choice of the CP phase 
$\delta$.  In the formulation of Eq.~(\ref{A13}) the sign of $A_{13}$ can be such that 
$\theta_{13}$ can assume negative values during the course of evolution
and in such situations one will have to talk about the evolution of 
$|\theta_{13}|$.
Our formulation in terms of 
$\theta_{13}^2$, as shown in Eq.~(\ref{th13sq-dot}), 
naturally avoids this problem.} 
of $\theta_{13}$ is avoided.
In terms of the new parameters $J_{\rm CP}$ and $ J_{\rm CP}^\prime$, 
the RG evolution equations for $\theta_{13}^2$ becomes
\beqa
\dot{{\theta_{13}^2} } &=& 
A_{13}^{sq} 
+  {\cal O}(\theta_{13}^2) \; , \label {th13sq-dot} \\
A^{sq}_{13} &=& 8 C y_\tau^2  \frac{ m_3}{\dma(1+\zeta)}  \Bigl\{  J_{\rm CP}
\left[ m_1 \sin{2 \phi_1} - (1+\zeta) m_2 \sin{2 \phi_2}\right] \Bigr. \nn \\
&& \Bigl. + J_{\rm CP}'
\left[ m_1 \cos{2 \phi_1} - (1+\zeta) m_2 \cos{2 \phi_2} - \zeta m_3\right]  
\Bigr\} \label{A13sq} \; .
\eeqa 
Thus the evolution equations in basis ${\cal P}_J$ are all non-singular 
and continuous at every point. 
In particular, even when $\delta$ shows a discontinuity,
$J_{\rm CP}$ as well as $J_{\rm CP}^\prime$ change in a continuous manner. 
This very fact can be used to write down the approximated integrated 
evolution equations for $J_{\rm CP}$, $J_{\rm CP}^\prime$ as
\beqa
J_{\rm CP}  &=& J_{\rm CP}^0 + k_{J_{\rm CP}} \Delta_\tau 
+ {\cal O}(\Delta_\tau \theta_{13}, \Delta_\tau^2) \; ,
\label{Jcp-int} \\
J_{\rm CP}^\prime &=& J_{\rm CP}^{\prime 0} + k_{J_{\rm CP}^\prime} \Delta_\tau 
+ {\cal O}(\Delta_\tau \theta_{13}, \Delta_\tau^2) \; ,
\label{JcpPrime-int}
\eeqa
where $J_{\rm CP}^0$, $J_{\rm CP}^{\prime 0}$ are the initial values at $\mu_0$ 
and $k_{J_{\rm CP}}$, $k_{J_{\rm CP}^\prime}$ can be obtained from Eqs.(\ref{AJ})-(\ref{AJprime}).
From the $J_{\rm CP}$, $J_{\rm CP}^\prime$ values, the Dirac CP phase $\delta$
can be determined unambiguously at any energy scale.



\section{RG evolution of neutrino masses and mixing in high energy seesaw models}
\label{sec:RG_full}

As discussed in Sec~\ref{sec:RG_effective}, the RG evolution of the 
neutrino masses and mixing parameters in the low energy effective theory 
is the same for all three types of seesaw scenarios 
and depends only on whether the low energy effective theory is 
the SM or the MSSM. However, this is not 
true in the high energy theory, when the heavy particles responsible for seesaw 
remain coupled to the theory. Hence in this case the RG evolution of the 
different neutrino parameters should also depend on the interaction of these 
heavy particles with other fields.  
The importance of including the effects from energy ranges above and between these
mass thresholds when analyzing RG effects in GUT models has been pointed out in 
\cite{tanimoto,Casas:1999tp,casas-deg,Mei:2004rn,King:2000hk,antusch-threshold-1,
antusch-threshold,antusch-LMA,Miura:2003if}. 
These effects are typically at least as important as the RG evolution effects from
below the thresholds since the relevant couplings may also be of 
order one.

The diagrams contributing to the renormalization constants of the 
different quantities at high energy are shown in 
\cite{kersten-thesis}, \cite{Schmidt-thesis} and \cite{triplet-paper} 
in case of Type-I, Type-II and Type-III seesaw respectively\footnote{
\cite{Schmidt-TypeII,Schmidt-thesis} actually considered Type-I + Type-II case where 
they have one heavy right-handed fermion and a triplet Higgs 
added to the SM (and the MSSM also). However, we will consider the 
three types of seesaw scenarios separately.
}. 
Finally the RG evolution equations for the charged lepton Yukawa matrix $Y_e$ and 
the heavy particle Yukawa matrix $Y_X$ and 
(here $X = {\rm N}$ for Type-I, $X = \Delta$ for Type-II and 
$X = \Sigma$ for Type-III) can be given as
\beqa
16 \pi^2 \beta_{Y_e} &=& Y_e F + \alpha_e Y_e \; , 
\label{Ye-RG-tot} \\
16 \pi^2 \beta_{Y_X} &=& \left\{ \barr{ll} 
Y_X G + \alpha_X Y_X & \quad {\small \text {Type-I \& III}} \; , \\
Y_X G + G^T Y_X + \alpha_X Y_X & \quad {\small \text {Type-II}} \; ,
\earr \right. 
\label{Yx-RG-tot} 
\eeqa
where $F$ and $G$ are defined as
\beqa
F &=& D_e Y_e^\dagger Y_e + D_X Y_X^\dagger Y_X \; , \label{F} \\
G &=& B_e Y_e^\dagger Y_e + B_X Y_X^\dagger Y_X \; . \label{G} 
\eeqa  
The quantities $D_e, D_X$, $B_e, B_X$ and $\alpha_e$, $\alpha_X$ 
with the SM as the low energy 
effective theory are given in Table~\ref{tab:complete_running}. 
The $\beta$-function for the heavy particle mass ${\mathbbm M}_X$ is given as
\beqa
16 \pi^2 \beta_{{\mathbbm M}_X} &=& \left\{ \barr{ll}
{\mathbbm M}_X \left( Y_X Y_X^\dagger \right)^T 
+ \left( Y_X Y_X^\dagger \right) {\mathbbm M}_X + 
\alpha_{{\mathbbm M}_X} {\mathbbm M}_X & {\small \text {Type-I \& III}} \; , \\
\alpha^\prime_{{\mathbbm M}_X} {\mathbbm M}_X^{-1} + \alpha_{{\mathbbm M}_X} {\mathbbm M}_X 
& {\small \text {Type-II}} \; , 
\earr \right. \nn \\
\label{Mx-RG-tot} 
\eeqa
with $\alpha_{{\mathbbm M}_X}$ and $\alpha^\prime_{{\mathbbm M}_X}$ defined by
\beqa
\alpha_{{\mathbbm M}_X} &=& \left\{ \barr{ll}
0 & {\small \text {Type-I}} \; , \\
4 \Lambda_1 + \Lambda_2 + \Tr[ Y_\Delta^\dagger Y_\Delta] 
- \frac{9}{5} g_1^2 - 6 g_2^2 & {\small \text {Type-II}} \; , \\
-12 g_2^2 & {\small \text {Type-III}} \; ,
\earr \right. \\
\alpha^\prime_{{\mathbbm M}_X} &=& 2 \Lambda_4 m^2 + \frac{1}{2} |\Lambda_6|^2 
\quad ({\small \text {Type-II}}) \; ,
\eeqa
where $m$ is the bare mass of the SM Higgs $\phi$ and $\Lambda_i$s are the 
couplings associated with the triplet Higgs $\Delta$, as given in Eq.~(\ref{LDelta_phi}). 

After electroweak symmetry breaking, these coupled heavy fields 
will contribute to the generation of the light neutrino mass via 
seesaw, as given in Sec~\ref{sec:typeI}-\ref{sec:typeIII} for 
the three different seesaw scenarios
\beqa
{\mathbbm m}_\nu = \left\{ \barr{ll}
-\frac{v^2}{2} Y_{\rm N}^T {\mathbbm M}_{\rm N}^{-1} Y_{\rm N} & \quad {\text{Type-I}} \; , \\
+\frac{v^2}{2} \frac{Y_{\Delta} \Lambda_6}{{\mathbbm M}_{\Delta}^2} & \quad {\text{Type-II}} \; , \\
-\frac{v^2}{2} Y_{\Sigma}^T {\mathbbm M}_{\Sigma}^{-1} Y_{\Sigma} & \quad {\text{Type-III}} \; .
\earr \right. 
\label{mnu-1}
\eeqa
In this section we would like to study the radiative corrections 
to the quantity 
\beqa 
Q \equiv \left\{ \barr{ll}
 Y_{\rm N}^T {\mathbbm M}_{\rm N}^{-1} Y_{\rm N} & \quad {\text{Type-I}} \; , \\
- \frac{Y_{\Delta} \Lambda_6}{{\mathbbm M}_{\Delta}^2} & \quad {\text{Type-II}} \; , \\
 Y_{\Sigma}^T {\mathbbm M}_{\Sigma}^{-1} Y_{\Sigma} & \quad {\text{Type-III}} \; ,
\earr \right. 
\label{Q-def}
\eeqa
which is a well-defined quantity at different energy scales to denote the 
contribution to the light neutrino mass matrix from the coupled heavy fields 
and finally gives the light neutrino mass matrix as 
${\mathbbm m}_\nu = -\frac{v^2}{2} Q$ after spontaneous symmetry breaking.

In the complete theory when the heavy fields are coupled, the RG evolution of $Q$ 
can be obtained from the running of $Y_X$ and ${\mathbbm M}_X$ (running of $\Lambda_6$ 
is also required in case of Type-II seesaw, which can be obtained in 
\cite{Schmidt-TypeII,zhang-TypeII}) and finally the evolution of $Q$ can be written as
\beqa
16 \pi^2 \beta_Q &=& Q P_Q + P_Q^T Q + \alpha_Q P_Q \; ,
\label{Q-RG-tot}
\eeqa 
with $P_Q$ defined as
\beq
P_Q = C^\prime_e Y_e^\dagger Y_e + C^\prime_X Y_X^\dagger Y_X \; .
\eeq
The quantities $C^\prime_e, C^\prime_X$ and $\alpha_Q$ for three 
seesaw scenarios with the SM as the low energy 
effective theory are also given in Table~\ref{tab:complete_running}. 
Finally ${\mathbbm m}_\nu$ can be obtained from $Q$ as 
${\mathbbm m}_\nu = -\frac{v^2}{2} Q$, after spontaneous symmetry breaking.

\begin{table}[t!] 
\begin{center} 
\small{
\begin{tabular}{|l|r|r|c|}
\hline
 & $D_e \;$ & $D_X \;$ & $\alpha_e$ \\
\hline
Type-I & $3/2$ & $-3/2$ & $T_I - \frac{9}{4} g_1^2 -\frac{9}{4} g_2^2$ \\
Type-II & $3/2$ & $3/2$ & $T_{II} - \frac{9}{4} g_1^2 -\frac{9}{4} g_2^2$ \\
Type-III & $3/2$ & $15/2$ & $T_{III} - \frac{9}{4} g_1^2 -\frac{9}{4} g_2^2$ \\
\hline
\hline
 & $B_e \;$ & $B_X \;$ & $\alpha_x$ \\
\hline
Type-I & $-3/2$ & $3/2$ & $T_I - \frac{9}{20} g_1^2 -\frac{9}{4} g_2^2$ \\
Type-II & $1/2$ & $3/2$ & $T_{II}^\prime - \frac{9}{10} g_1^2 -\frac{9}{2} g_2^2$ \\
Type-{III} & $5/2$ & $5/2$ & $T_{III} - \frac{9}{20} g_1^2 - \frac{33}{4} g_2^2$ \\
\hline
\hline
 & $C^\prime_e \;$ & $C^\prime_X \;$ & $\alpha_Q$ \\
\hline
Type-I & $-3/2$ & $1/2$ & $2 T_I - \frac{9}{10} g_1^2 - \frac{9}{2} g_2^2$ \\
Type-II & $1/2$  & $3/2$ & $T_{II} - 2 T_{II}^\prime - 3g_2^2 + \lambda + f(\Lambda_i)$ \\
Type-III & $5/2$  & $3/2$ & $2 T_{III} - \frac{9}{10} g_1^2 - \frac{9}{2} g_2^2$ \\
\hline
\hline
 & $C_e \;$ & $C_X \;$ & $\alpha_\kappa$ \\
\hline
Type-I & $-3/2$ & $1/2$ & $2 T_I + \lambda - 3 g_2^2$ \\
Type-II & $-3/2$  & $3/2$ & $2 T_{II} + \lambda - 3 g_2^2$  \\
Type-III & $-3/2$ & $3/2$ & $2 T_{III} + \lambda - 3 g_2^2$ \\
\hline
\end{tabular}
}
\end{center}
\caption{The quantities defining the running of $Y_e$, $Y_X$, $Q$ and $\kappa$ 
for three types of seesaw scenarios \cite{antusch-CP,Schmidt-TypeII,triplet-paper}. 
Here $T_I \equiv \Tr [Y_e^\dagger Y_e + Y_{\rm N}^\dagger Y_{\rm N} + 3 Y_u^\dagger Y_u + 3 Y_d^\dagger Y_d]$, 
$T_{II} \equiv \Tr[Y_e^\dagger Y_e + 3 Y_u^\dagger Y_u + 3 Y_d^\dagger Y_d ]$,  
$T^\prime_{II} \equiv \Tr[ Y_\Delta^\dagger Y_\Delta]$ and 
$T_{III} \equiv \Tr[ Y_e^\dagger Y_e + 3 Y_\Sigma^\dagger Y_\Sigma + 3 Y_u^\dagger Y_u + 3 Y_d^\dagger Y_d]$. 
The function $f(\Lambda_i)$ is defined as $f(\Lambda_i) = -8 \Lambda_1 - 2 \Lambda_2 
- 4 \Lambda_4 + 8 \Lambda_5 - \left( 4 \Lambda_4 m^2 + |\Lambda_6|^2 \right) {\mathbbm M}_\Delta^{-2}$, 
where $m$ is the bare mass of the SM Higgs doublet $\phi$ and $\Lambda_i$ are the couplings 
associated with the triplet Higgs $\Delta$, as defined in Eq.~(\ref{LDelta_phi}).
\label{tab:complete_running}}
\end{table}

In case of Type-I seesaw, the heavy fermion singlets do not have any gauge 
interactions and hence the presence of these singlets does not affect the running 
of the gauge couplings. However, for the other two seesaw scenarios, the RG 
evolution of the gauge couplings $g_1$, $g_2$, $g_3$ will depend on the 
number of heavy fields present. Let $n$ be the number of heavy fields 
present at some energy scale. Then the evolution of the gauge couplings 
can be written as 
\beq
16 \pi^2 \beta_{g_i} = b_i g_i^3 \; ,
\label{gauge-run}
\eeq
where the values for $b_i$s in the three seesaw scenarios 
are tabulated in Table~\ref{tab:gauge}. 
\begin{table}[t!] 
\begin{center} 
\small{
\begin{tabular}{|l|c|c|c|}
\hline
 & $b_1$ & $b_2$ & $b_3$ \\
\hline
SM & $41/10$ & $-19/6$ & $-7$ \\
Type-I & $41/10$ & $-19/6$ & $-7$   \\
Type-II & $41/10 + 3n/5$ & $-19/6 + 2n/3$  &  $-7$ \\
Type-III & $41/10$ &$-19/6 + 4n/3$  &  $-7$ \\
\hline
\end{tabular}
}
\end{center}
\caption{$b_i$ in three types of seesaw scenarios with the SM as the 
low energy theory\cite{antusch-CP,zhang-TypeII,triplet-paper}. 
Here $n$ is the number of heavy particles coupled to the theory at 
any particular energy scale.
\label{tab:gauge}}
\end{table}
As given in the Table~\ref{tab:gauge}, none of the heavy fields has any 
strong interactions and hence $b_3$ is always the same as its SM value. 
The Higgs triplets present in Type-II seesaw have $Y=1$ and thus couple 
to both U(1)$_Y$ and SU(2)$_L$ gauge fields, while the triplet fermions 
in Type-III seesaw have $Y=0$ and thus couple only to the SU(2)$_L$ gauge 
fields.

We do not give the evolution equations for the different Higgs couplings 
or for the up- and down-type quark Yukawa couplings here. The evolution 
equations for these couplings  
can be obtained in \cite{antusch-CP,Schmidt-TypeII,triplet-paper} for the 
different seesaw scenarios.

\subsection{Sequential decoupling of heavy fields}

The most general case of the high energy theories would have been the one 
when there are any arbitrary number of right-handed singlets in Type-I 
seesaw, or any number of triplet scalars in case of Type-II seesaw or 
arbitrary number of fermion triplets in Type-III seesaw and these 
heavy fields decouple one by one at different thresholds.

Let us first consider the most general case of Type-I and Type-III seesaw 
when there are $r$ heavy fields (singlets in case of Type-I and triplets in case of Type-III) 
having masses $M_1 < M_2 < \cdots < M_{r-1} < M_r$. 
We consider a quantity ${\mathbbm R}$ which contains the 
contribution from the coupled as well as decoupled heavy fields 
at any energy scale and is also well-defined at all $\mu$, and then 
finally gives the light neutrino mass matrix as
\beqa
{\mathbbm m}_\nu = -\frac{v^2}{2} {\mathbbm R} \; ,
\eeqa 
after electroweak symmetry breaking.

Above the heaviest  mass $M_r$, all the $r$-fields are coupled to 
the theory and contribute to ${\mathbbm R}$ as
\beqa 
\accentset{(r+1)}{{\mathbbm R}} &=& \accentset{(r+1)}{Q} \; , 
\eeqa
where $\accentset{(r+1)}{Q}$ denotes the contribution from the 
$r$ coupled heavy fields and is given by
\beqa 
\accentset{(r+1)}{Q} = \accentset{(r+1)}{Y}_{X}^T \; ~\accentset{(r+1)}{\mathbbm M}_{X}^{-1} 
\;~ \accentset{(r+1)}{Y}_{X} ~~~~~~~ (\mu > M_r) \; , 
\label{Q-rplus1}
\eeqa
where $X \equiv {\rm N}$ for Type-I and $X \equiv {\Sigma}$ for Type-III seesaw. 
Here $\accentset{(r+1)}Y_X$ is a $[r \times n_F]$ dimensional matrix  
($n_F$ is the number of flavors, which is 3 in our case) given as
\beqa 
\accentset{(r+1)}{Y}_X =  
\left( \begin{array}{ccc} 
{(y_X)}_{1,1} & \cdots & {(y_X)}_{1,n_F} \\ 
\vdots & & \vdots \\ 
{(y_X)}_{r,1} & \cdots & {(y_X)}_{r,n_F}\\ 
\end{array} \right) \;. 
\label{Y-rplus1} 
\eeqa   
$\accentset{(r+1)}{{\mathbbm M}_X}$ is a $[r \times r]$ matrix and 
$\accentset{(r+1)}{Q}$ as well as $\accentset{(r+1)}{{\mathbbm R}}$ 
is a $[n_F \times n_F]$ dimensional matrix. 
We use the super-indices just to keep track of the number of coupled fields.   
Below the scale $M_r$, the heaviest of the heavy fields decouples from the  
theory. Integrating out this degree of freedom gives rise to an effective 
operator $\accentset{(r)}\kappa$.   
The matching condition at $\mu = M_r$ is  
\beqa 
\left. {\accentset{(r)}\kappa}_{ij} \right\arrowvert_{M_r}&=& 
\left. 2 {(\accentset{(r+1)}{Y_X^T})}_{ir}\; 
(M_r)^{-1} \;  
{(\accentset{(r+1)} {Y_X})}_{rj} \right\arrowvert_{M_r} \; , 
\label{mnum} 
\eeqa 
where no summation over `$r$' is implied and $i,j \in \{ 1,2,\cdots,n_F \}$. 
This condition ensures the continuity of ${\mathbbm R}$ at 
$\mu=M_r$. In order to get the value of the threshold $M_r$, we need to  
write the above matching condition in the basis where  
${\mathbbm M}_X = {\rm Diag}( M_1, M_2, \cdots , M_r)$. 
Here it is worth mentioning that the matching scale has to be found  
carefully since ${\mathbbm M}_X$ itself runs with the energy scale, 
{\it i.e.} $M_i = M_i(\mu)$. The threshold scale $M_i$ is therefore to 
be understood as $M_i(\mu=M_i)$.

In the energy range $M_{r-1} < \mu < M_r$, ${\mathbbm R}$ will be given as 
\beqa 
\accentset{(r)}{{\mathbbm R}} &=&  
\frac{1}{2}\accentset{(r)}{\kappa} +  
\accentset{(r)}{Q} \; . 
\label{mnum-1} 
\eeqa 
The first term in Eq.~(\ref{mnum-1}) is the contribution  
of the  integrated out heavy fermion of mass $M_r$ through the  
effective operator $\accentset{(r)}{\kappa}$. The second term 
represents the contribution of the remaining $(r-1)$ heavy fermions, 
which are still coupled to the theory. $\accentset{(r)}{{\mathbbm M}}_X$ is 
now a $[(r-1) \times (r-1)]$ matrix while $\accentset{(r)}{Y}_X$ is a 
$[(r-1) \times n_F]$ dimensional matrix given as 
\beqa 
{Y}_X \rightarrow  
\left( \begin{array}{ccc} 
{(y_X)}_{1,1} & \cdots & {(y_X)}_{1,n_F} \\ 
\vdots & & \vdots \\ 
{(y_X)}_{r-1,1} & \cdots & {(y_X)}_{r-1,n_F}\\ 
\hline 
0 & \cdots & 0 
\end{array} \right)  \: 
\begin{array}{cl} 
\left.\begin{array}{c}\\[1.4cm]\end{array}\right\} 
& =\accentset{(r)}{Y}_X \;, 
 \\ 
\left.\begin{array}{c}\\[0.1cm]\end{array}\right\} 
& \begin{array}{l} M_r \; \text{integrated out}\;.\end{array} 
\end{array} \; 
\label{YX-r} 
\eeqa 
Finally $\accentset{(r)}{\mathbbm M}_X$ and $\accentset{(r)}{Y}_X$ 
constitute $\accentset{(r)}{Q}$, which is $[n_F \times n_F]$ dimensional.   
The matching condition at $\mu = M_{r-1}$ is  
\beqa 
\left. {\accentset{(r-1)}\kappa}_{ij} \; \;\right\arrowvert_{M_{r-1}} & = &  
\left. {\accentset{(r)}\kappa}_{ij} \; \right\arrowvert_{M_{r-1}} +  
\left. 2 {(\accentset{(r)}{Y}_X^{\; T})}_{i(r-1)} \;  
(M_{r-1})^{-1} \; 
{(\accentset{(r)}{Y}_X)}_{(r-1)j} \right\arrowvert_{M_{r-1}} \; , \quad
\label{matching} 
\eeqa 
where no summation over `$(r-1)$' is to be taken.

Generalizing the above sequence, we can say that if we consider  
the intermediate energy region between the $(n-1)^{\rm th}$ and the  
$n^{\rm th}$ threshold, i.e.  
$M_n > \mu > M_{n-1}$, then  all the heavy fields from masses $M_r$   
down to $M_n$ have been decoupled.
In this region the Yukawa matrix $\accentset{(n)}{Y}_X$  
will be $[(n-1) \times n_F]$ dimensional that couples the
$(n-1)$ coupled fields with $n_F$ flavors and will be given as 
\beqa 
{Y}_X \rightarrow  
\left( \begin{array}{ccc} 
{(y_X)}_{1,1} & \cdots & {(y_X)}_{1,n_F} \\ 
\vdots & & \vdots \\ 
{(y_X)}_{n-1,1} & \cdots & {(y_X)}_{n-1,n_F}\\ 
\hline 
0 & \cdots & 0 \\ 
\vdots &  & \vdots \\ 
0 & \cdots & 0 
\end{array} \right)  \: 
\begin{array}{cl} 
\left.\begin{array}{c}\\[1.7cm]\end{array}\right\} 
& =\accentset{(n)}{Y}_X\;, 
 \\ 
\left.\begin{array}{c}\\[1.0cm]\end{array}\right\} 
& \begin{array}{l}\text{heavy fermions with }\\
\text{masses} \; M_n \mbox{---} M_r \\ 
\text{integrated out}\;.\end{array} 
\end{array} \; 
\label{YX-n} 
\eeqa 
$\accentset{(n)}{\mathbbm M}_X$ will be $[(n-1) \times (n-1)]$ dimensional 
matrix involving the mass terms of all the coupled heavy fields. 
In this energy range ${\mathbbm R}$ will be  
\beqa 
\accentset{(n)}{{\mathbbm R}} &=&  
\frac{1}{2}\accentset{(n)}{\kappa} + 
 \accentset{(n)}{Q} \; , 
\label{mnu} 
\eeqa 
with 
\beq 
\Qn \equiv \accentset{(n)}{Y}_X^{\; T} \;~ \accentset{(n)}{{\mathbbm M}}_X^{-1} \;~\accentset{(n)}{Y}_X \; . 
\label{Qn} 
\eeq 
Note that ${\mathbbm R}$, $\kappa$ and $Q$ are $[n_F \times n_F]$ matrices. 
The matching condition  
at $\mu = M_{n}$ is given by Eq.~(\ref{matching}) with  
$r$ replaced by $(n+1)$. 

At low energies $\mu < M_1$, when all the heavy fields are
decoupled, $Q(\mu) = \accentset{(1)}{Q}(\mu) = 0$ and 
$\accentset{(1)}{\mathbbm R}(\mu) = (1/2) \accentset{(1)}{\kappa}(\mu)$.
\begin{figure} 
\begin{center} 
\includegraphics[scale=0.4]{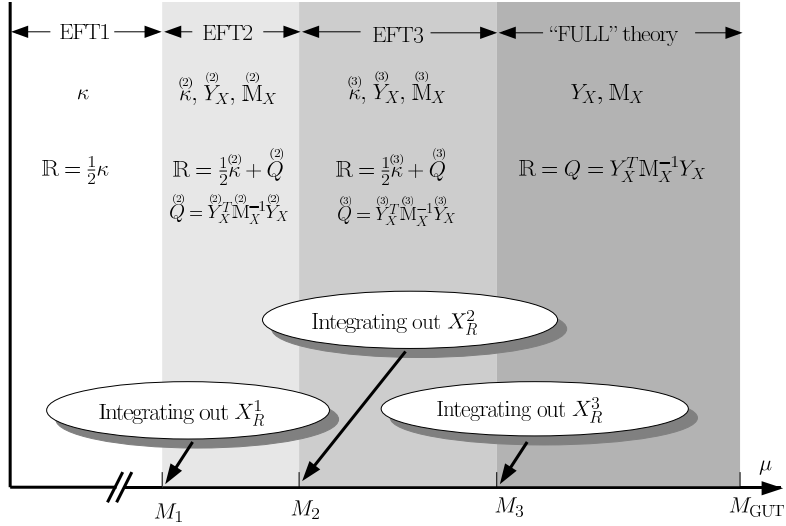} 
\caption{Sequential decoupling of the heavy fermions 
and construction of ${\mathbbm R}$ at different energy scales, 
for $r=3$. Here $X \equiv {\rm N}$ in case of Type-I and $X \equiv \Sigma$ for Type-III seesaw. 
Finally the light neutrino mass matrix is given as 
${\mathbbm m}_\nu \equiv -\frac{v^2}{2} {\mathbbm R}$ after 
spontaneous symmetry breaking, where $v$ is the vacuum expectation 
value of the SM Higgs.
\label{sequancial-decouple-fig}} 
\end{center} 
\end{figure} 
Fig.~\ref{sequancial-decouple-fig} shows the expressions for 
${\mathbbm R}$ at different energy scales for the case of three heavy fermions 
{\it i.e.} for $r=3$.
Finally the light neutrino mass matrix ${\mathbbm m}_\nu$ 
is obtained as ${\mathbbm m}_\nu \equiv -\frac{v^2}{2} {\mathbbm R}$, after 
the electroweak symmetry breaking.

In case of Type-I and Type-III seesaw, the concept of sequential decoupling is 
important since atleast two heavy fields are needed to generate the observed 
pattern of the light neutrino mass matrix and the heavy fermions can have non-degenerate 
masses in general. However, the case of Type-II seesaw is much simpler since 
only one heavy scalar triplet is sufficient to give rise to the small masses of 
the three active neutrinos and their mixings.
Hence in this case one has only one threshold at $\mu = {\mathbbm M}_\Delta$ and 
${\mathbbm R}$ at the two different energy regimes will be given by
\beqa
{\mathbbm R} = \left\{  \barr{ll}
Q & (\mu > {\mathbbm M}_\Delta) \; ,\\
\frac{1}{2} \kappa & (\mu < {\mathbbm M}_\Delta) \; ,\\
\earr \right.
\eeqa 
where $Q$ and $\kappa$ are defined in Eq.~(\ref{Q-def}) and Table~\ref{seesaw-summary} 
respectively. The matching conditions at $\mu = {\mathbbm M}_\Delta$ will be given by
\beqa 
\left. \kappa_{ij} \; \;\right\arrowvert_{{\mathbbm M}_\Delta} & = &  
- \frac{2}{{\mathbbm M}_\Delta^2} \left.\Lambda_6  {(Y_\Delta)}_{ij} \right\arrowvert_{{\mathbbm M}_\Delta} \; , \quad
\label{matching-TypeII} \\
\left. \lambda \; \;\right\arrowvert_{{\mathbbm M}_\Delta} &=& 
\left. \lambda \; \;\right\arrowvert_{{\mathbbm M}_\Delta} + 
\frac{2}{{\mathbbm M}_\Delta^2} \left. |\Lambda_6|^2 \; \;\right\arrowvert_{{\mathbbm M}_\Delta} \; .
\label{matching-lambda}
\eeqa 
Finally the light neutrino mass matrix ${\mathbbm m}_\nu$ 
is obtained as ${\mathbbm m}_\nu \equiv -\frac{v^2}{2} {\mathbbm R}$, after 
the electroweak symmetry breaking.

For the sake of convenience and to be consistent 
with the existing literature, we will refer to 
${\mathbbm m}_\nu = -\frac{v^2}{2} {\mathbbm R}$ as the effective 
light neutrino mass matrix at any energy scale, in the rest of the 
thesis. However, it must be understood that the quantity `$v$' is 
present only after electroweak symmetry breaking and hence, 
strictly speaking, this relation is valid only in that energy regime, 
while ${\mathbbm R}$ is a well-defined quantity at all energy scales.


Finally, the RG evolution of the light neutrino mass matrix will be given by
\beq
16 \pi^2 \beta_{{\mathbbm m}_\nu} = P^T {\mathbbm m}_\nu + {\mathbbm m}_\nu P 
+ \alpha_\nu {\mathbbm m}_\nu \; ,
\label{RG-mnu-tot}
\eeq
which is of the same form as Eq.~(\ref{beta-mnu}). Here 
$P \equiv P_Q$, $\alpha_\nu \equiv \alpha_Q$ for $\mu > M_3$ and 
$P \equiv P_\kappa$, $\alpha_\nu \equiv \alpha_\kappa$ for $\mu < M_1$. 
In case of Type-I and Type-III seesaw, the running of the 
light neutrino masses in between the thresholds will be given by the 
running of both $Q$ and $\kappa$, as shown in Eq.~(\ref{mnu}). 
For the energy scale $M_n > \mu > M_{n-1}$, the relevant quantities 
will be $\Qn$ and $\accentset{(n)}{\kappa}$. The RG evolution of $\Qn$ 
can be obtained from Table~\ref{tab:complete_running} with 
the substitution $Y_X \to \accentset{(n)}{Y}_X$ and 
${\mathbbm M}_X \to \accentset{(n)}{\mathbbm M}_X$. In case of Type-II 
seesaw there is only one threshold and the running of all relevant 
quantities contributing to the running of the neutrino mass matrix 
can be read off from Table~\ref{tab:complete_running}.

It can be seen from the interaction of the heavy fields that they do not 
contribute to the 1-loop correction of the effective operator 
$\accentset{(n)}{\kappa}$, even when they are coupled to the theory. So 
the evolution of the effective vertex $\accentset{(n)}{\kappa}$ can 
be written as
\beqa
16 \pi^2 \beta_{\; \accentset{(n)}\kappa } 
&=& P_\kappa^T \; \accentset{(n)}{\kappa} + \accentset{(n)}{\kappa} P_\kappa 
+ \alpha_\kappa \accentset{(n)}{\kappa}
\label{beta-kappan}
\eeqa
where
\beqa
P_\kappa &=& C_e Y_e^\dagger Y_e 
+ C_X  \accentset{(n)}{Y}_X^\dagger \accentset{(n)}{Y}_X  \; , 
\label{P-nu}
\eeqa
and the quantities $C_e$, $C_X$ and $\alpha_\kappa$ can be obtained from 
Table~\ref{tab:complete_running} and are independent of $n$, the number 
of heavy fields coupled at any energy scale.


\subsection{RG evolution of neutrino mixing parameters}

In order to evaluate the RG evolution of the light neutrino masses and the 
mixing parameters from the matrix evolutions, we proceed in the 
same way as done in Sec~\ref{subsec:effective_parameter_RG}. 
Without the lose of generality, we choose to work in the basis in which 
$\accentset{(n)}{\mathbbm M}_X$ and $Y_e^\dagger Y_e$ are diagonal at the 
high energy.

As stated in the last section, 
for $\mu > M_3$ and $\mu < M_1$, the evolutions of ${\mathbbm m}_\nu$ 
will be given by the evolution of $Q$ ($P = P_Q$) and 
$\kappa$ ($P = P_\kappa$) respectively and hence the evolution 
of the angles, phases and light neutrino 
masses can be given in simple analytic forms. 
$P$ and $F$, in Eqs.(\ref{RG-mnu-tot}) and (\ref{Ye-RG-tot}), 
are $3 \times 3$ matrices with the rows and columns  
representing generations. We denote the elements of $P$ and $F$  
by $P_{fg}$ and $F_{fg}$. If we write the 
evolution equations in the basis ${\cal P}_\delta$, the apparent singularity 
at $\theta_{13} \to 0$ will be present, as can be seen from 
\cite{Schmidt-thesis,antusch-threshold-1}. As already discussed in 
Sec~\ref{sec:subtlety}, this singularity can be removed using the basis 
${\cal P}_J$ \cite{triplet-paper}. Hence we discuss the RG evolution of the 
mixing angles, phases and the light neutrino masses in the ${\cal P}_J$ 
basis in the following sections.

\subsubsection{Evolution of mixing angles}

\begin{table}[t!] 
\centering 
  \begin{tabular}{|l|c|c|} 
\hline 
 & $32\pi^2 \, \Dot\theta_{12}$  
  & $32\pi^2 \, \Dot\theta_{23}$ \\ 
\hline 
 
 $ P_{11}$ 
 & $\mathcal{Q}^+_{12}\sin 2\theta_{12}$ 
  & $0$ \\ 
 
   $P_{22}$ 
 & $-\mathcal{Q}^+_{12}\sin 2\theta_{12}c_{23}^2$ 
  &  $\left( \mathcal{Q}^+_{23}c_{12}^2  
+ \mathcal{Q}^+_{13}s_{12}^2 \right) \sin 2\theta_{23}$  
\\ 
 
 $P_{33}$ 
 & $-\mathcal{Q}^+_{12} \sin 2\theta_{12}s_{23}^2$ 
  & $-\!\left( \mathcal{Q}^+_{23}c_{12}^2 + \mathcal{Q}^+_{13}s_{12}^2 \right)  
\sin 2\theta_{23}$ \\ 
 
 $\re P_{21}$ 
 & $2\mathcal{Q}^+_{12}\cos 2\theta_{12} c_{23}$ 
 & $\left( \mathcal{Q}^+_{23} - \mathcal{Q}^+_{13} \right)  
\sin 2\theta_{12}s_{23}$ \\ 
 
 $\re P_{31}$ 
 & $-2\mathcal{Q}^+_{12} \cos 2\theta_{12} s_{23}$ 
 & $\left( \mathcal{Q}^+_{23} - \mathcal{Q}^+_{13} \right)\sin 2\theta_{12} 
 c_{23}$ \\ 
  
 $\re P_{32}$ 
 & $\mathcal{Q}^+_{12} \sin 2\theta_{12}\sin 2\theta_{23}$ 
 & $2\! \left( \mathcal{Q}^+_{23}c_{12}^2  
 + \mathcal{Q}^+_{13}s_{12}^2 \right) \cos 
 2\theta_{23}$ \\ 
  
 $\im P_{21}$ 
 & $4 \mathcal{S}_{12}c_{23}$ 
 & $2 \left( \mathcal{S}_{23} - \mathcal{S}_{13} \right) \sin 2\theta_{12}s_{23} $ 
 \\ 
  
 $\im P_{31}$ 
 & $-4\mathcal{S}_{12} s_{23}$ 
 & $2 \left( \mathcal{S}_{23} - \mathcal{S}_{13} \right)  
\sin 2\theta_{12}c_{23} $ 
 \\ 
  
 $\im P_{32}$ 
 & $0$ 
 & $4 \left( \mathcal{S}_{23}c_{12}^2  
+ \mathcal{S}_{13}s_{12}^2 \right)$\\
\hline 
\end{tabular}
\begin{tabular}{|l|c|} 
\hline 
 & $64\pi^2 \, \Dot{\overline{\theta_{13}^2}}$ 
\\ 
\hline 
 $ P_{11}$ &  $0$  \\ 
 $P_{22}$ &  $\left( \mathcal{\widetilde{A}}^+_{23} - 
\mathcal{\widetilde{A}}^+_{13} \right)  
\sin 2\theta_{12}\sin 2\theta_{23}$  \\ 
 $P_{33}$ &  $-\left( \mathcal{\widetilde{A}}^+_{23}  
- \mathcal{\widetilde{A}}^+_{13} \right) \sin 2\theta_{12} \sin 2\theta_{23}$ \\ 
 $\re P_{21}$  & $4 \left( \mathcal{\widetilde{A}}^+_{13}c_{12}^2  
+ \mathcal{\widetilde{A}}^+_{23}s_{12}^2 \right)  s_{23}$  \\ 
 $\re P_{31}$  & $4 \left( \mathcal{\widetilde{A}}^+_{13}c_{12}^2  
+ \mathcal{\widetilde{A}}^+_{23}s_{12}^2 
 \right)c_{23}$ \\ 
 $\re P_{32}$  & $2\! \left( \mathcal{\widetilde{A}}^+_{23}  
- \mathcal{\widetilde{A}}^+_{13} \right)  \sin 2\theta_{12}\cos 2\theta_{23}$ \\ 
 $\im P_{21}$  & $4\left( \mathcal{\widetilde{B}}^-_{13}c_{12}^2  
+ \mathcal{\widetilde{B}}^-_{23}s_{12}^2 \right) s_{23}$ \\ 
 $\im P_{31}$  & $4 \left( \mathcal{\widetilde{B}}^-_{13}c_{12}^2  
+ \mathcal{\widetilde{B}}^-_{23}s_{12}^2 \right)c_{23}$ \\ 
 $\im P_{32}$  &  $2 \left( \mathcal{\widetilde{B}}^-_{23}  
- \mathcal{\widetilde{B}}^-_{13} \right) \sin 2\theta_{12}$ \\
\hline 
\end{tabular} 
\caption{Coefficients of $P_{fg}$ in the RG evolution  
equations of the mixing angles 
$\theta_{12}$, $\theta_{13}^2$ 
and $\theta_{23}$,   
in the limit $\theta_{13}\to 0$ \cite{antusch-threshold-1,triplet-paper}.  
\label{tab-theta12-theta23} } 
\end{table}
%
%
Running of the two large mixing angles  
$\theta_{12}$ and $\theta_{23}$ in the basis ${\cal P}_J$, as given in  
Table~\ref{tab-theta12-theta23}, is also the same as that in the  
${\cal P}_\delta$ basis since the quantities  
$\mathcal{S}_{ij}$ and $\mathcal{Q}_{ij}^\pm$, defined as 
\beqa 
\barr{lll}
\mathcal{Q}^\pm_{13} = \frac{|m_3 \pm m_1 e^{2 i \phi_1}|^2}{\Delta 
  m^2_\mathrm{atm}\left(1+\zeta\right)} \; , &
\mathcal{Q}^\pm_{23} = \frac{|m_3 \pm m_2 e^{2 i \phi_2}|^2}{\Delta 
  m^2_\mathrm{atm}}\; ,
\mathcal{Q}^\pm_{12} = \frac{|m_2 e^{ 2 i \phi_2} \pm m_1 
  e^{2 i \phi_1}|^2}{\dms} \; , 
\label{Qij}
\earr
\\
\barr{lll}
 \mathcal{S}_{13} = \frac{m_1 m_3 \sin{2 \phi_1}}{\Delta m^2_\mathrm{atm} 
\left( 1+\zeta \right)} \; , &
\mathcal{S}_{23} =  
\frac{m_2 m_3 \sin{ 2 \phi_2} }{\Delta m^2_\mathrm{atm}} \; , &
\mathcal{S}_{12} =  
\frac{m_1 m_2 \sin{( 2 \phi_1- 2 \phi_2)} }{\dms} \; , 
\label{Sij} 
\earr
\eeqa
%
depend on the mass eigenvalues and Majorana phases only, and 
not on the Dirac CP phase $\delta$. 
However the running of $\theta_{13}^2$, as seen from the  
Table~\ref{tab-theta12-theta23},  
depends on the quantities $\mathcal{\widetilde{A}}^\pm_{ij}$, 
$\mathcal{\widetilde{B}}^\pm_{ij}$ defined as
\beqa 
\mathcal{\widetilde{A}}^\pm_{13} =  
\frac{ 4 \left(m_1^2+m_3^2\right) J'_{\rm CP} \pm 8 m_1m_3  
( J'_{\rm CP} \cos{2 \phi_1} +  
 J_{\rm CP} \sin{2 \phi_1})}{  a \Delta m_\mathrm{atm}^2 
\left(1+\zeta\right)} \; , \label{Atilde13} \\ 
\mathcal{\widetilde{A}}^\pm_{23} =  
\frac{ 4 \left(m_2^2+m_3^2\right) J'_{\rm CP} \pm 8 m_2m_3  
( J'_{\rm CP} \cos{2 \phi_2} +  
 J_{\rm CP} \sin{2 \phi_2})}{ a \Delta m_\mathrm{atm}^2} \; ,  
\label{Atilde23} 
\eeqa 
\beqa 
\mathcal{\widetilde{B}}^\pm_{13} =  
\frac{ 4 \left(m_1^2+m_3^2\right) J_{\rm CP} \pm 8 m_1m_3  
( J_{\rm CP} \cos{2 \phi_1} -  
 J'_{\rm CP} \sin{2 \phi_1})}{  a \Delta m_\mathrm{atm}^2 
\left(1+\zeta\right)} \; , \label{Btilde13}\\ 
\mathcal{\widetilde{B}}^\pm_{23} =  
\frac{ 4 \left(m_2^2+m_3^2\right) J_{\rm CP}  
\pm 8 m_2m_3 ( J_{\rm CP} \cos{2 \phi_2} -  
 J'_{\rm CP} \sin{2 \phi_2})}{  a \Delta m_\mathrm{atm}^2} \; , 
\label{Btilde23} 
\eeqa 
where $a \equiv s_{12} c_{12} s_{23} c_{23}$.
Clearly these quantities depend on $J_{\rm CP}$, $J'_{\rm CP}$ 
in addition to the masses and Majorana phases and hence are basis-dependent. 
In the ${\cal P}_J$ basis, all the the quantities appearing in the  
evolution equations (\ref{Atilde13}) -- (\ref{Btilde23})  
have finite well-defined limits for $\theta_{13} \to 0$ and 
so will be $\theta_{13}^2$ at any energy scale.
\begin{table}[htbp]
\centering
\begin{tabular}{|l|c|c|c|c|c|c|}
\hline
&\multicolumn{3}{|c|}{$\Dot\theta_{12}$}
&\multicolumn{3}{|c|}{$\Dot\theta_{23}$}\\\hline
& d. & n.h. &i.h. &d. & n.h. &i.h. \\\hline
$P_{11}$ & $\frac{m^2}{\dms}$ & $1$ & $\zeta^{-1}$
&$\mathcal{O}(\theta_{13})$ & $\mathcal{O}(\theta_{13})$ & $\mathcal{O}(\theta_{13})$
\\
$P_{22}$ & $\frac{m^2}{\dms}$ & $1$ & $\zeta^{-1}$
&$\frac{m^2}{\Delta m^2_\text{atm}}$ & $1$ & $1$
\\
$P_{33}$ & $\frac{m^2}{\dms}$ & $1$ & $\zeta^{-1}$
&$\frac{m^2}{\Delta m^2_\text{atm}}$ & $1$ & $1$
\\
$\re P_{21}$ & $\frac{m^2}{\dms}$ & $1$ & $\zeta^{-1}$
&$\frac{m^2}{\Delta m^2_\text{atm}}$ & $\sqrt{\zeta}$ & $\mathcal{O}(\theta_{13})$
\\
$\re P_{31}$ & $\frac{m^2}{\dms}$ & $1$ & $\zeta^{-1}$
&$\frac{m^2}{\Delta m^2_\text{atm}}$ & $\sqrt{\zeta}$ & $\mathcal{O}(\theta_{13})$
\\
$\re P_{32}$ & $\frac{m^2}{\dms}$ & $1$ & $\zeta^{-1}$
&$\frac{m^2}{\Delta m^2_\text{atm}}$ & $1$ & $1$
\\
$\im P_{21}$ & $\frac{m^2}{\dms}$ & $\mathcal{O}(\theta_{13})$ & $\zeta^{-1}$
&$\frac{m^2}{\Delta m^2_\text{atm}}$ & $\sqrt{\zeta}$ & $\mathcal{O}(\theta_{13})$
\\
$\im P_{31}$ & $\frac{m^2}{\dms}$ & $\mathcal{O}(\theta_{13})$ & $\zeta^{-1}$
&$\frac{m^2}{\Delta m^2_\text{atm}}$ & $\sqrt{\zeta}$ & $\mathcal{O}(\theta_{13})$
\\
$\im P_{32}$ &  $\mathcal{O}(\theta_{13})$ & $\mathcal{O}(\theta_{13})$ & $\mathcal{O}(\theta_{13})$
&$\frac{m^2}{\Delta m^2_\text{atm}}$ & $\sqrt{\zeta}$ & $\mathcal{O}(\theta_{13})$
\\ \hline
\end{tabular}
\begin{tabular}{|l|c|c|c|}
\hline
&\multicolumn{3}{|c|}{$\Dot\theta_{13}^2$}\\\hline
& d. & n.h. & i.h.\\\hline
$P_{11}$ & ${\cal O}(\theta_{13}^2)$ &${\cal O}(\theta_{13}^2)$  & ${\cal O}(\theta_{13}^2)$\\
$P_{22}$ & $\frac{m^2}{\dma}\theta_{13}$ & $\sqrt{\zeta} \theta_{13}$ & ${\cal O}(\theta_{13}^2)$\\
$P_{33}$ & $\frac{m^2}{\dma}\theta_{13}$ & $\sqrt{\zeta} \theta_{13}$ & ${\cal O}(\theta_{13}^2)$\\
$\re P_{21}$ & $\frac{m^2}{\dma}\theta_{13}$ & $\theta_{13}$ & $\theta_{13}$\\
$\re P_{31}$ & $\frac{m^2}{\dma}\theta_{13}$ & $\theta_{13}$  & $\theta_{13}$\\
$\re P_{32}$ & $\frac{m^2}{\dma}\theta_{13}$ & $\sqrt{\zeta} \theta_{13}$  & ${\cal O}(\theta_{13}^2)$\\
$\im P_{21}$ & $\frac{m^2}{\dma}\theta_{13}$ & $\theta_{13}$ & $\theta_{13}$\\
$\im P_{31}$ & $\frac{m^2}{\dma}\theta_{13}$ & $\theta_{13}$  & $\theta_{13}$\\
$\im P_{32}$ & $\frac{m^2}{\dma}\theta_{13}$ & $\sqrt{\zeta} \theta_{13}$  & ${\cal O}(\theta_{13}^2)$\\
\hline
\end{tabular}
\caption{Generic enhancement and suppression factors for the evolution
 of the angles, yielding an estimate of the size of the RG effect \cite{antusch-threshold-1}. 
The table entries correspond to the terms in the mixing parameter RG evolution 
equations with the coefficient given by the first column.
 A `$1$' indicates that there is no generic enhancement or 
suppression. `d.' stands for a degenerate neutrino mass spectrum, {\it i.e.} 
$\Delta m_\mathrm{atm}^2\ll m_1^2\sim m_2^2\sim m_3^2\sim m^2$. `n.h.' denotes a 
normally hierarchical spectrum, {\it i.e.} $m_1\ll m_2\ll m_3$, and 
`i.h.' means an inverted hierarchy, {\it i.e.} $m_3\ll m_1\lesssim m_2$.}
\label{tab:EnhancementFactorsAngles}
\end{table}

Table~\ref{tab:EnhancementFactorsAngles} shows the generic enhancement and 
suppression factors \cite{antusch-threshold-1} for the evolution of the 
mixing angles, which is useful to estimate the RG evolution effects on the 
angles, when the active neutrinos are quasi-degenerate 
($\Delta m_\mathrm{atm}^2\ll m_1^2\sim m_2^2\sim m_3^2\sim m^2$ and this case is 
denoted by `d.'), or have normal mass hierarchy ($m_1\ll m_2\ll m_3$ and denoted by `n.h.') 
or inverted mass hierarchy ($m_3\ll m_1\lesssim m_2$ and denoted by `i.h.').
From Table~\ref{tab:EnhancementFactorsAngles} we see that all terms in 
$\Dot\theta_{12}$ are enlarged by  $m^2/\dms$ for
quasi-degenerate masses.  Thus, there will be large RG effects, if the different
terms do not cancel each other.  The term involving $\im P_{32}$ is an
exception, because its leading order is proportional to $\theta_{13}$, so that
it only plays a role in special cases. Also the terms involving $\im P_{21}$ and 
$\im P_{31}$ will have small contributions for small values of $(2\phi_1 - 2\phi_2)$.
In the case of a strong normal hierarchy, there is no enhancement. 
For an inverted hierarchy, where the evolution is generically enhanced by 
$\zeta^{-1}$, because the masses $m_1$ and $m_2$ are almost degenerate.

Both for $\theta_{23}$ and $\theta_{13}^2$, the evolution does not depend on 
$P_{11}$ for $\theta_{13}=0$. For these two mixing angles, the enhancement and 
suppression factors are similar (for $\theta_{13}^2$ evolution, there is always 
an extra factor of $\theta_{13}$ compared to $\theta_{23}$, as expected). 
The terms proportional to the other $P_{fg}$ are enhanced by
$m^2/\dma$ in the degenerate case, so that effects are expected to be 
significant, but smaller than $\theta_{12}$ running. 
For both hierarchical spectra, the running is slow, as can be seen from 
Table~\ref{tab:EnhancementFactorsAngles}. In case of diagonal $P$ (or with 
$P_{32}$ as the only non-zero off-diagonal entry) and 
inverted hierarchy, there will be no running for $\theta_{13}^2$ if 
$\theta_{13}=0$. However, this is no longer true if $P_{21}$ or $P_{31}$ is non-zero.

Thus in the evolution equations of the mixing angles, the generic 
characteristics of the terms which are proportional to the diagonal 
elements of $P$ in the high energy theory is the same as those in the 
low energy effective theory, as already discussed in 
Eqs.~(\ref{A12})--(\ref{A13}) in Sec~\ref{sec:RG-mixingparam_effective}.

If the diagonal elements are equal, their contributions to the RG evolution 
equations cancel exactly. This follows from the fact that the mixing angles do 
not change under RG evolution, if $P$ is the identity matrix and thus does not 
distinguish between the flavors. As can be seen in the next few sections, 
this statement holds also for the RG evolution of the CP phases. 
It provides a consistency check for the results. 
Interesting new effects occur for non-zero off-diagonal elements in $P$. Some of
their coefficients in the evolution equations do not vanish for vanishing mixings, 
{\it e.g.} the coefficient of $P_{21}$ in $\Dot \theta_{12}$ in 
Table~\ref{tab-theta12-theta23}, and thus non-zero mixing angles are 
generated radiatively. This is in striking contrast to the region below the
see-saw scale, as can be checked from Eqs.~(\ref{A12})--(\ref{A13}).


\subsubsection{Evolution of $J_{\rm CP}, J_{\rm CP}^\prime$}

\begin{table}[ht!] 
\centering 
\begin{tabular}{|l|c|c|}
\hline 
& $ 64\pi^2 \,  \Dot J_{\rm CP} / a$ 
& $ 64\pi^2 \,  \Dot J'_{\rm CP} / a $  
\\\hline
  
$P_{11}$ 
& $0$  
& $0$  
\\ 
 
$P_{22}$ 
& $-4 a \calG_s^-$ 
& $2 a ( \calG_0^- - 2 \calG_c^-)$ 
\\ 
  
$P_{33}$ 
& $4 a \calG_s^-$ 
& $ - 2 a ( \calG_0^- -  2 \calG_c^-)$  
\\ 
 
$\re P_{21}$ 
& $4 s_{23} \calG_s^+$ 
& $2 s_{23} (\calG_0^+ + 2 \calG_c^+)$  
\\ 
 
$\re P_{31}$ 
& $4 c_{23} \calG_s^+$ 
& $ 2 c_{23} (\calG_0^+ + 2 \calG_c^+)$  
\\ 
 
$\re P_{32}$ 
& $ - 2 \sin{2 \theta_{12}} \cos{2 \theta_{23}} \, \calG_s^- \, $  
& $ \sin{2 \theta_{12}} \cos{2 \theta_{23}} (\calG_0^- - 2 \calG_c^-)$  
\\ 
 
$\im P_{21}$ 
& $ 2 s_{23} (\calG_0^+ - 2 \calG_c^+)$ 
& $ 4 s_{23} \calG_s^+ $  
\\ 
 
$\im P_{31}$ 
&  $ 2 c_{23} (\calG_0^+ - 2 \calG_c^+)$ 
& $ 4 c_{23} \calG_s^+ $  
\\ 
 
$\im P_{32}$ 
& $  \sin{2 \theta_{12}} (\calG_0^- + 2 \calG_c^-)$ 
& $ - 2 \sin{2 \theta_{12}} \calG_s^- $  
\\\hline 
\end{tabular} 
\caption{
Coefficients of $P_{fg}$ in the RG evolution equations of  
the Jarlskog invariant $J_{\rm CP}$, the quantity $J'_{\rm CP} \equiv
J_{\rm CP} \cot \delta$, in the limit $\theta_{13}\to 0$. The convention
used here is $a \equiv s_{12} c_{12} s_{23} c_{23}$, and 
$J_{\rm CP} \equiv (a/2) s_{13} c_{13}^2 \sin \delta$ \cite{triplet-paper}.  
\label{tab-J} } 
\end{table}

The coefficients for the RG evolution of $J_{CP}$ and $J'_{CP}$  
are presented in Table~\ref{tab-J}, where the quantities  
$\calG_{0,c,s}^{\pm}$ are given by 
\beqa 
\calG_0^\pm &=& \frac{m_2^2 + m_3^2}{\dmsq_\mathrm{atm}} \pm   
\frac{m_1^2 + m_3^2}{\dmsq_\mathrm{atm} (1+\zeta)} \; , \label{G0}\\ 
\calG_s^\pm &=& \frac{m_1 m_3 \sin{2 \phi_1}}{\dmsq_\mathrm{atm} (1+\zeta)} \pm   
\frac{m_2 m_3 \sin{2 \phi_2}}{\dmsq_\mathrm{atm}} \; , \label{Gs}\\ 
\calG_c^\pm &=& \frac{m_1 m_3 \cos{2 \phi_1}}{\dmsq_\mathrm{atm} (1+\zeta)} \pm   
\frac{m_2 m_3 \cos{2 \phi_2}}{\dmsq_\mathrm{atm}} \;\label{Gc} . 
\eeqa 
Thus the quantities defined in Eqs.~(\ref{G0})--(\ref{Gc}) are 
functions of masses and Majorana phases and hence are well-defined at all 
energies and at every point in the parameter space. Thus 
Table~\ref{tab-J} shows that the running of $J_{\rm CP}$, $J'_{\rm CP}$ 
does not depend on themselves and hence independent of the Dirac CP 
phase $\delta$ upto ${\cal O}(\theta_{13}^0)$. It also shows that if 
$P$ is identity (or proportional to identity), there will be 
no RG evolution, as expected.
%

\begin{table}[htbp]
\centering
\begin{tabular}{|l|c|c|c|c|c|c|}
\hline
&\multicolumn{3}{|c|}{$\Dot J_{\rm CP}/a$}
&\multicolumn{3}{|c|}{$\Dot J_{\rm CP}^\prime /a$}\\\hline
& d. & n.h. &i.h. &d. & n.h. &i.h. \\\hline
$P_{11}$ &  $\mathcal{O}(\theta_{13})$ & $\mathcal{O}(\theta_{13})$  &  $\mathcal{O}(\theta_{13})$
&$\mathcal{O}(\theta_{13})$ & $\mathcal{O}(\theta_{13})$ & $\mathcal{O}(\theta_{13})$
\\
$P_{22}$ & $\frac{m^2}{\dma}$ & $\sqrt{\zeta}$ & $\mathcal{O}(\theta_{13})$
&$\frac{m^2}{\Delta m^2_\text{atm}}$ & $\zeta$ & $\zeta$
\\
$P_{33}$ & $\frac{m^2}{\dma}$ & $\sqrt{\zeta}$ & $\mathcal{O}(\theta_{13})$
&$\frac{m^2}{\Delta m^2_\text{atm}}$ & $\zeta$ & $\zeta$
\\
$\re P_{21}$ & $\frac{m^2}{\dma}$ & $\sqrt{\zeta}$ & $\mathcal{O}(\theta_{13})$
&$\frac{m^2}{\Delta m^2_\text{atm}}$ & $1$ & $1$
\\
$\re P_{31}$ & $\frac{m^2}{\dma}$ & $\sqrt{\zeta}$ & $\mathcal{O}(\theta_{13})$
&$\frac{m^2}{\Delta m^2_\text{atm}}$ & $1$ & $1$
\\
$\re P_{32}$ & $\frac{m^2}{\dma}$ & $\sqrt{\zeta}$ & $\mathcal{O}(\theta_{13})$
&$\frac{m^2}{\Delta m^2_\text{atm}}$ & $\zeta$ & $\zeta$
\\
$\im P_{21}$ & $\frac{m^2}{\dma}$ & $1$ & $1$
&$\frac{m^2}{\Delta m^2_\text{atm}}$ & $\sqrt{\zeta}$ & $\mathcal{O}(\theta_{13})$
\\
$\im P_{31}$ & $\frac{m^2}{\dma}$ & $1$ & $1$
&$\frac{m^2}{\Delta m^2_\text{atm}}$ & $\sqrt{\zeta}$ & $\mathcal{O}(\theta_{13})$
\\
$\im P_{32}$ & $\frac{m^2}{\dma}$  & $\zeta$ & $\zeta$
&$\frac{m^2}{\Delta m^2_\text{atm}}$ & $\sqrt{\zeta}$ & $\mathcal{O}(\theta_{13})$
\\ \hline
\end{tabular}
\caption{Generic enhancement and suppression factors for the evolution
 of $J_{\rm CP}$ and $J^\prime_{\rm CP}$, yielding an estimate of the size 
of the RG effect. 
\label{tab:EnhancementFactorsJcp} }
\end{table}

From the generic enhancement and suppression factors for the RG evolution 
of $J_{\rm CP}$ and $J_{\rm CP}^\prime$ given in 
Table~\ref{tab:EnhancementFactorsJcp} it can be seen that for degenerate 
light neutrino masses the coefficients are enlarged by the factor $m^2/\dma$, 
for all $P_{fg}$ except $P_{11}$. The leading contribution from $P_{11}$ 
comes only at ${\cal O}(\theta_{13})$ and is also independent of the mass 
ordering of the neutrinos. For $\Dot J_{\rm CP}$, the contributions from 
the other two diagonal elements $P_{22}$ and $P_{33}$ are suppressed by $\sqrt\zeta$ 
for normal hierarchy, while for inverted hierarchy the leading contribution 
is only at ${\cal O}(\theta_{13})$. For $\Dot J^\prime_{\rm CP}$ the 
evolution is suppressed by $\zeta$ in both the cases.

From Table~\ref{tab-J} it can be seen that even if $\theta_{13}$ is zero to 
start with so that $J_{\rm CP} = J_{\rm CP}^\prime = 0$ at the high scale, it can 
be generated radiatively. This is true even with a diagonal $P$, if $P$ is 
not proportional to identity. 
This happens in the low energy effective theory also, as can be seen from 
Eqs.~(\ref{AJ})--(\ref{AJprime}). However, if the diagonal elements are equal, their 
contributions cancel exactly and there will be no running at all. 
Table~\ref{tab-J} also suggests that for 
non-zero off-diagonal elements of $P$, $J_{\rm CP}$ and $J_{\rm CP}^\prime$ can be 
generated radiatively even when all the mixing angles are zero at the high scale 
and this is very different from what is expected in the effective theory below the 
seesaw scale.


\subsubsection{Evolution of Majorana phases}

The expressions for the running of the Majorana phases are
the same in ${\cal P}_J$ and ${\cal P}_\delta$. 
Table~\ref{tab-phi} shows the running of the difference between 
the Majorana phases $|\phi_1 - \phi_2|$ \cite{antusch-threshold-1,triplet-paper}.
%
\begin{table}[t!] 
\centering 
\begin{tabular}{|l|c|}
\hline 
& $32 \pi^2 (\dot\phi_1 - \dot\phi_2)$ 
\\\hline
  
$P_{11}$  & $-4\mathcal{S}_{12}\cos 2\theta_{12}$ 
\\ 
$P_{22}$  & $4\mathcal{S}_{12}c_{23}^2 \cos 2\theta_{12}$ 
\\ 
  
$P_{33}$ & $4 \mathcal{S}_{12}s_{23}^2\cos 2\theta_{12}$
\\ 
 
$\re P_{21}$ & $-8\mathcal{S}_{12}c_{23}\cos 2\theta_{12}\cot 2\theta_{12}$ 

\\ 
 
$\re P_{31}$ & $8\mathcal{S}_{12}s_{23}\cos 2\theta_{12} \cot 2\theta_{12}$ 
\\ 
 
$\re P_{32}$ & $-4\mathcal{S}_{12} \cos 2\theta_{12}\sin 2\theta_{23}$ 
\\ 
 
$\im P_{21}$ & $-4\mathcal{Q}^-_{12}c_{23} \cot 2\theta_{12}$ 
\\ 
 
$\im P_{31}$ & $4\mathcal{Q}^-_{12} s_{23}\cot 2\theta_{12}$ 
\\ 
 
$\im P_{32}$ & 0
\\\hline 
\end{tabular} 
\caption{
Coefficients of $P_{fg}$ in the RG evolution equations of  
the Majorana phase difference $(\phi_1-\phi_2)$, 
in the limit $\theta_{13}\to 0$ \cite{antusch-threshold-1,triplet-paper}. 
\label{tab-phi} } 
\end{table}
\begin{table}[h!]
\centering
\begin{tabular}{|l|c|c|c|}
\hline
&\multicolumn{3}{|c|}{$\Dot\phi_1 - \Dot\phi_2$}\\\hline
& d. & n.h. & i.h.\\\hline
$P_{11}$ & $\frac{m^2}{\dms}$ & ${\cal O}(\theta_{13})$ & $\zeta^{-1}$ \\
$P_{22}$ & $\frac{m^2}{\dms}$ & ${\cal O}(\theta_{13})$  &$\zeta^{-1}$  \\
$P_{33}$ & $\frac{m^2}{\dms}$ & ${\cal O}(\theta_{13})$ & $\zeta^{-1}$ \\
$\re P_{21}$ & $\frac{m^2}{\dms}$ & ${\cal O}(\theta_{13})$ & $\zeta^{-1}$ \\
$\re P_{31}$ & $\frac{m^2}{\dms}$ & ${\cal O}(\theta_{13})$  & $\zeta^{-1}$ \\
$\re P_{32}$ & $\frac{m^2}{\dms}$ & ${\cal O}(\theta_{13})$  & $\zeta^{-1}$ \\
$\im P_{21}$ & $\frac{m^2}{\dms}$ & $1$ & $\zeta^{-1}$ \\
$\im P_{31}$ & $\frac{m^2}{\dms}$ & $1$  & $\zeta^{-1}$ \\
$\im P_{32}$ & ${\cal O}(\theta_{13})$ & ${\cal O}(\theta_{13})$ & ${\cal O}(\theta_{13})$\\
\hline
\end{tabular}
\caption{Generic enhancement and suppression factors for the evolution
 of the difference of Majorana phases $(\phi_1 - \phi_2)$ \cite{antusch-threshold-1}.}
\label{tab:EnhancementFactorsMajorana}
\end{table}

As can be seen from Table~\ref{tab:EnhancementFactorsMajorana}, the generic 
enhancement factors for the RG evolution of $(\phi_1 - \phi_2)$ are very similar 
to those for the running of $\theta_{12}$, for degenerate light neutrino masses or 
for an inverted hierarchy. For normal hierarchy, there is no running if $P$ 
is real, 
upto the zeroth order of $\theta_{13}$, which implies that each of 
the Majorana phases runs by equal amount. The running of individual Majorana 
phases is discussed in \cite{antusch-threshold-1}. The 
running of the Majorana phases is also important to understand the 
evolution of the mixing angles, since all the quantities defined in 
Eqs.~(\ref{Qij})--(\ref{Btilde23}) depend on the Majorana phases. RG 
evolution of the Majorana phases controls the running of $J_{\rm CP}$, 
$J_{\rm CP}^\prime$ also.


\subsubsection{Evolution of light neutrino masses}

Table~\ref{tab-mi} shows the RG evolution of the light neutrino masses. 
As can be seen, the coefficients are independent of $J_{\rm CP}$, 
$J_{\rm CP}^\prime$ and hence the expressions remain the same in the basis $P_\delta$.

\begin{table}[t!] 
\centering 
\begin{tabular}{|l|c|c|c|}
\hline 
& $16 \pi^2 \dot{m_1}/m_1$ 
& $16 \pi^2 \dot{m_2}/m_2$ 
& $16 \pi^2 \dot{m_3}/m_3$ \\\hline
$\alpha_\nu$ 
& $1$ 
& $1$ 
& $1$ \\
$P_{11}$ 
& $2 c_{12}^2$ 
& $2 s_{12}^2$ 
& $0$ \\
$P_{22}$ 
& $2 s_{12}^2 c_{23}^2$ 
& $2 c_{12}^2 c_{23}^2$ 
& $2 s_{23}^2$ \\
$P_{33}$ 
& $2 s_{12}^2 s_{23}^2$ 
& $2 c_{12}^2 s_{23}^2$ 
& $2 c_{23}^2$ \\
$\re P_{21}$ 
& $-2 \sin{2\theta_{12}}c_{23}$ 
& $2 \sin{2\theta_{12}}c_{23}$ 
& $0$ \\
$\re P_{31}$ 
& $2 \sin{2\theta_{12}}s_{23}$ 
& $-2 \sin{2\theta_{12}}s_{23}$ 
& $0$ \\
$\re P_{32}$ 
& $-2 \sin{2\theta_{23}}s_{12}^2$ 
& $-2 \sin{2\theta_{23}}c_{12}^2$ 
& $2 \sin{2\theta_{23}}$ \\
$\im P_{21}$ 
& $0$ 
& $0$ 
& $0$ \\
$\im P_{31}$ 
& $0$ 
& $0$ 
& $0$ \\
$\im P_{32}$ 
& $0$ 
& $0$ 
& $0$ \\\hline
\end{tabular} 
\caption{
Coefficients of $P_{fg}$ in the RG evolution equations of  
the neutrino masses $m_i$ $\{ i=1,2,3\}$,  
in the limit $\theta_{13}\to 0$ \cite{antusch-threshold-1,triplet-paper}.
\label{tab-mi} } 
\end{table}

As is clear from Table~\ref{tab-mi}, the evolution of $m_i$ is proportional to 
itself. This is a general characteristic of the running of the mass eigenvalues at 
all energy scales. As a consequence, the
mass eigenvalues can never run from a finite value to zero or vice versa. 
However, this conclusion is very specific to the 1-loop running of the masses, and 
breaks down when the 2-loop contributions are taken into account \cite{ishidori}.

As already discussed in Sec~\ref{sec:RG-mass_effective}, below the 
see-saw scales, the evolution of the mass eigenvalues is, to a 
good approximation, described by a universal scaling caused by the
flavor-independent part of the RG evolution equations proportional to $\alpha_\nu$. 
This flavor-independent term becomes smaller at high energies. Therefore, the 
flavor-dependent terms play a more important role above the see-saw scales. The 
importance of the flavor-dependent part increases if entries of $Y_X$ 
become of order one. Thus between and above the see-saw scales, 
the running may become strongly influenced by the Yukawa couplings of the 
heavy fields.

RG evolution of $\dms$ and $\dma$, the quantities important for neutrino 
oscillations, can be obtained using Table~\ref{tab-mi}, and is also 
discussed in \cite{antusch-threshold-1}.


\subsubsection{Contribution from $U_e$}
 
As already stated, we choose to work in the basis in which 
$\accentset{(n)}{\mathbbm M}_X$ is diagonal. 
Hence from the Eqs.~(\ref{kappaI}) 
and (\ref{kappaIII}) we get that $\accentset{(n)}{Y}_{\rm N}$ and 
$\accentset{(n)}{Y}_\Sigma$ will have non-zero off-diagonal 
components. So even if one starts with diagonal $Y_e$ 
({\it i.e.} $Y_e = {\rm Diag}(y_e,y_\mu,y_\tau)$) at the high scale,  
non-zero off-diagonal elements of $Y_e$ will be generated  
through Eqs.~(\ref{Ye-RG-tot}) and (\ref{F}) since  
$\accentset{(n)}{Y}_X^{\;\dagger} \accentset{(n)}{Y}_X$ 
is not diagonal. Thus the contribution from $U_e$ to $U_{\rm PMNS}$, 
as given in Eq.~(\ref{Upmns-Ue-Unu}), will be finite and there 
will be finite contribution to the running of masses and mixing 
above and between the thresholds through $F$ and $\alpha_e$. 
Since $\alpha_e$ is flavor diagonal, it will contribute to  
the running of $y_e$, $y_\mu$ and $y_\tau$,  
while off-diagonal components of $F$ will  
contribute additional terms in the $\beta$-functions of angles  
and phases. To evaluate the contributions from the off-diagonal 
components of $Y_e$, we consider the evolution of $U_e$ as 
\cite{antusch-threshold-1}
\beq
\frac{d U_e}{dt} = U_e X \; ,
\label{Ue-run}
\eeq
where $t \equiv \ln(\mu/{\rm GeV})/16 \pi^2$ and $X$ is an anti-Hermitian 
matrix which can be determined from Eqs.~(\ref{Ue-def}) and (\ref{Ye-RG-tot}) 
to have the form \cite{antusch-threshold-1}
\beq
16 \pi^2 X_{ij} = \frac{y_j^2 + y_i^2}{y_j^2 - y_i^2} \left( U_e^\dagger F U_e \right)_{ij} \quad (i \ne j)\; ,
\label{X-def}
\eeq
where $y_1 = y_e$ and so on. The diagonal parts of $X$, which only influence
the evolution of the unphysical phases, remain undetermined. 
Using Eqs.~(\ref{RG-U}) and (\ref{Ue-run}), 
one can write from Eq.~(\ref{Upmns-Ue-Unu})
\beq
\frac{d U_{\rm PMNS}}{dt} = U_{\rm PMNS} T + X^\dagger U_{\rm PMNS} \; .
\label{run-Upmns}
\eeq
%
\begin{table}[t!] 
\centering 
  \begin{tabular}{|l|c|c|c|} 
\hline 
& $ \; 16 \pi^2 \, \Dot \theta_{12}^{U_e} \; $  
& $ \; 16 \pi^2 \, {\Dot {\theta^2}_{13}}^{U_e} \; $ 
& $ \; 16 \pi^2 \, \Dot \theta_{23}^{U_e} \; $  
\\\hline 
 
$F_{11}$ 
& $0$ 
& $0$ 
& $0$ 
\\ 
 
$F_{22}$ 
& $0$ 
& $0$ 
& $0$ 
\\ 
 
$F_{33}$ 
& $0$ 
& $0$ 
& $0$ 
\\ 
 
$\re F_{21}$ 
& $-c_{23}$ 
& $ - 4 s_{23} J'_{\rm CP} / a $ 
& $0$ 
\\ 
 
$\re F_{31}$ 
& $s_{23}$ 
& $- 4 c_{23} J'_{\rm CP} / a $ 
& $0$ 

\\ 
 
$\re F_{32}$ 
& $0$ 
& $0$ 
& $1$ 
\\ 
 
$\im F_{21}$ 
& $0$ 
& $ - 4 s_{23} J_{\rm CP} / a $ 
& $0$ 

\\ 
 
$\im F_{31}$ 
& $0$ 
& $- 4 c_{23} J_{\rm CP} / a$ 
& $0$ 

\\ 
 
$\im F_{32}$ 
& $0$ 
& $0$ 
& $0$ 
\\\hline  
\end{tabular} 
\caption{Coefficients of $F_{fg}$ in the RG evolution equations of all 
the angles ($\theta_{12}$, $\theta_{13}^2$, $\theta_{23}$), 
in the limit $\theta_{13}\to 0$. The convention
used here is $a \equiv s_{12} c_{12} s_{23} c_{23}$, and 
$J_{\rm CP} \equiv (a/2) s_{13} c_{13}^2 \sin \delta$.  
We neglect $y_e$ and $y_\mu$ compared to $y_\tau$, 
and take vanishing flavor phases \cite{triplet-paper}.}
\label{tab-F-angle} 
\end{table}
\begin{table}[t!] 
\centering 
  \begin{tabular}{|l|c|c|c|c|} 
\hline 
& $ \; 16 \pi^2 \, \Dot J_{\rm CP}^{U_e} \; $ 
& $ \; 16 \pi^2 \, \Dot J_{\rm CP}^{' \; U_e} \; $ 
& $ \; 16 \pi^2 \, \Dot \phi_1^{U_e} \; $  
&  \; $16 \pi^2 \, \Dot \phi_2^{U_e} \; $ 
\\\hline 
 
$F_{11}$ 
& $0$ 
& $0$ 
& $0$ 
& $0$ 
\\ 
 
$F_{22}$ 
& $0$ 
& $0$ 
& $0$ 
& $0$ 
\\ 
 
$F_{33}$ 
& $0$ 
& $0$ 
& $0$ 
& $0$ 
\\ 
 
$\re F_{21}$ 
& $0$ 
& $- s_{23} a /2$ 
& $0$ 
& $0$ 
\\ 
 
$\re F_{31}$ 
& $0$ 
& $- c_{23} a / 2$ 
& $0$ 
& $0$  
\\ 
 
$\re F_{32}$ 
& $0$ 
& $0$ 
& $0$ 
& $0$  
\\ 
 
$\im F_{21}$ 
& $ - s_{23} a/2 \; $ 
& $0$ 
& $ \; c_{23} c_{12} / s_{12} \; \; $ 
& $ \; -c_{23} s_{12} / c_{12} \; \; $ 
\\ 
 
$\im F_{31}$ 
& $- c_{23} a / 2$ 
& $0$ 
& $ \; -s_{23} c_{12} / s_{12} \; \; $ 
& $ \; s_{23} s_{12} / c_{12} \; \; $ 
\\ 
 
$\im F_{32}$ 
& $0$ 
& $0$ 
&$ \; -1/(c_{23} s_{23}) \; \; $ 
&$ \; -1/(c_{23} s_{23}) \; \; $ 
\\\hline  
\end{tabular} 
\caption{Coefficients of $F_{fg}$ in the RG evolution equations of 
$J_{\rm CP}, J'_{\rm CP}$ and the Majorana phases $\phi_i$  
in the limit $\theta_{13}\to 0$ \cite{triplet-paper}.  } 
\label{tab-F-J-phi} 
\end{table}
%
Using the expressions for $T$ and $X$ from Eqs.~(\ref{def-ReT}), (\ref{def-ImT}) and 
(\ref{X-def}), one gets the coupled equations for the angles and phases 
from Eq.~(\ref{run-Upmns}). As suggested by Eq.~(\ref{run-Upmns}), the 
contributions from the first term are already tabulated in 
Tables~\ref{tab-theta12-theta23}, \ref{tab-J}, \ref{tab-phi} 
and \ref{tab-mi}, and the contribution from the second term 
is the additional contribution because of the 
off-diagonal entries in $Y_e^\dagger Y_e$ generated in course of 
RG evolution. These additional terms in the $\beta$-functions of angles  
and phases are tabulated in Tables~\ref{tab-F-angle} and 
\ref{tab-F-J-phi}, respectively. These contributions will just get added  
to the $P_{fg}$ contribution for the evolution of the quantities given 
in Tables~\ref{tab-theta12-theta23}, \ref{tab-J} and \ref{tab-phi}. 
Note that the $F_{fg}$ coefficients are $\lesssim {\cal O}(1)$,  
whereas the $P_{fg}$ coefficients are $\gtrsim {\cal O}(m_i^2/\dma)$.  
Since the running is significant only when $m_i^2 \gg \dma$, in  
almost all the region of interest $P_{fg}$ contributions  
dominate over the $F_{fg}$ contribution.\footnote{
If the running of the mixing angles $\theta_{12}$ and 
$\theta_{23}$ is large to make the angles close to zero or 
$\pi/2$ at some energy scale, the coefficients of some of the $\im F_{fg}$ may become 
large for $\dot \phi_i$, as can be seen from Table~\ref{tab-F-J-phi}.
}

In Type-II seesaw, we consider only one triplet Higgs, 
and so ${\mathbbm M}_\Delta$ is a number and hence from Eq.~(\ref{kappaII}) 
we see that $Y_\Delta$ may be chosen to be diagonal in general. 
Thus if $Y_e$ is chosen to be diagonal at 
high energy, it will remain so and the same procedure as in 
Sec~\ref{subsec:effective_parameter_RG} can be followed with the 
change that here the running of $Q$ is to be considered instead of $\kappa$.

Note that the analytical expressions obtained in Eq.~(\ref{Qij}) onwards, 
and those given in the tables, are valid only in the two extreme regions 
$\mu>M_3$ and $\mu < M_1$. For the intermediate energy scales, 
${\mathbbm m}_\nu$ will receive contributions from both $\accentset{(n)}{\kappa}$  
and $\Qn$. In the SM these two quantities have non-identical evolutions, 
as seen from Eqs.~(\ref{beta-kappan}) and (\ref{Q-RG-tot}), and therefore the net  
evolution of $Y_e$ and ${\mathbbm m}_\nu$ is rather complicated and needs 
numerical studies.

Quantitative studies have been made to show that the threshold effects may 
have dramatic consequences and can make many high energy neutrino mass models 
compatible with the current oscillation data at low energy which would have 
been excluded otherwise and vice versa. To illustrate the fact, bimaximal 
mixing scenario ($\theta_{12} = \theta_{23} = \pi/4$, $\theta_{13} = 0$) \cite{bimax} 
is not allowed by the current oscillation experiment data, as can be seen from 
Table~\ref{tab:bounds}. But it is possible to make this symmetry allowed at the 
high energy when threshold effects are taken into account, in case of both Type-I 
\cite{antusch-LMA} and Type-III \cite{triplet-paper} seesaw. 



\section{Conclusions}

In the framework of the standard model (SM) of particle physics, neutrinos 
are massless at the tree level as well as at loop level. Hence one has to 
extend the SM in order to explain the tiny active neutrino masses observed 
experimentally. The most favored mechanisms to generate such small neutrino 
masses are the seesaw mechanisms, in which small 
active neutrino masses are generated at some high energy scale.

All these models predict the light neutrino masses and the mixing 
parameters at some high energy which corresponds, in some way, to the mass scale 
of the new fields added to the SM to generate the light neutrino masses. 
But since the experimental data are available at the laboratory energy scale, 
one needs to include the effects of renormalization group (RG) evolution. 
Unlike the quark sector where RG evolutions are quite small 
because of the hierarchical quark masses and small mixings, the effect 
of RG evolution on the neutrino masses and the mixing parameters are important. 
In this paper we reviewed the seesaw mechanisms that generate the light neutrino 
masses and the RG evolution of the neutrino masses and mixing parameters 
in the seesaw scenarios.

The low energy effective operator to generate 
the light neutrino masses is the same for all the three types of seesaws and 
thus the RG evolution of the parameters depend only on the effective theory 
{\it i.e.} whether it is the SM or the Minimal Supersymmetric Standard Model (MSSM), 
which has been discussed in detail. 
The Pontecorvo-Maki-Nakagawa-Sakata (PMNS) parametrization of the neutrino 
mixing matrix is characterized by the fact that at $\theta_{13} = 0$, 
the Dirac CP phase $\delta$ is unphysical. This leads to the singular 
behavior of $\dot \delta$ at $\theta_{13} = 0$.
However, this singularity is unphysical, 
since all the elements of the neutrino mixing matrix $U_{\rm PMNS}$ are continuous
at $\theta_{13} = 0$, and in fact the value of $\delta$ there should be immaterial.
The singularity also creeps in the running of $\theta_{13}$, while evolution of the 
other parameters is well-behaved. However, as discussed here, it is possible to 
express the RG evolution of all the parameters as continuous first-order differential 
equations if one chooses the basis to be 
${\cal P}_J = \{m_i, \theta_{12}, \theta_{23}, \theta_{13}^2, \phi_i, J_{\rm CP}, J'_{\rm CP} \}$, 
instead of the conventional 
${\cal P}_\delta \equiv \{m_i, \theta_{ij}, \phi_i, \delta \}$ 
basis. The evolution equations in the new ${\cal P}_J$ basis and their 
approximate integrated forms have also been discussed.

When the RG evolution in the high energy theory is considered, one needs to take 
the effects of the heavy fields into account carefully, since the evolution will 
depend on their interactions with the other fields. Moreover, the heavy fields 
will decouple from the theory step by step at their respective mass scales 
and start contributing through the effective operators, so the threshold 
effects are to be considered and matching conditions are to be imposed. 
We considered the evolution in three seesaw scenarios and with the SM as the 
effective theory. For energy regimes higher than the mass of the 
heaviest particle and lower than the lightest one, the evolution can be expressed 
by simple analytic formulae and qualitative understanding of the RG evolution is 
possible independent of the low energy theory considered. 
However, the final evolution of any parameter will depend on the choice of the 
high energy seesaw scenario, as well as the low energy effective theory.

RG evolution can have dramatic effects on the masses and the mixing parameters 
in the neutrino sector, especially when threshold effects are taken into account. 
These effects can make many high energy neutrino mass models compatible with 
the current oscillation data at low energy which would have been excluded otherwise 
and vice verse. 
Since precision data is expected from the upcoming neutrino experiments, 
it is important to consider the RG evolution effects while 
talking about the neutrino mass models, and it then be possible exclude different 
classes of models and to gather knowledge about the possible high scale symmetries.


\section*{Acknowledgement}

S.R. would like to thank Prof. Amol Dighe for his useful suggestions, guidance 
and comments on the final manuscript.
The work was partially supported by the 
Max Planck -- India Partnergroup project between Tata Institute  
of Fundamental Research and Max Planck Institute for Physics.

\section*{Appendix: Diagonalization of neutrino mass matrix}
\label{appendix}
\addcontentsline{toc}{section}{Appendix: Diagonalization of neutrino mass matrix}

\renewcommand{\theequation}{A.\arabic{equation}}
\setcounter{equation}{0}

To check the diagonalization procedure, let us consider the 
case when there are arbitrary `$n$' number of heavy right-handed neutrinos and 
three active neutrino species. Then the neutrino mass 
matrix ${\cal M}_\nu$ is a $n \times n$ matrix given by 
Eq.~(\ref{mass-matrix-nu}) as
\beqa
{\cal M}_\nu = \left( \barr{cc} 0 & {\mathbbm m}_D \\ 
{\mathbbm m}_D^T & {\mathbbm M}_{\rm N} \earr \right) \; ,
\eeqa
where ${\mathbbm m}_D$ is a $3 \times n$ matrix, and ${\mathbbm M}_{\rm N}$ 
is $n \times n$. Let us now consider the unitary transformation 
\beqa
{\left( \barr{cc} A & B \\ C & D \earr \right)}^\dagger
\left( \barr{cc} 0 & {\mathbbm m}_D \\ {\mathbbm m}_D^T & {\mathbbm M}_{\rm N} \earr \right)
{\left( \barr{cc} A & B \\ C & D \earr \right)}^\ast
= 
\left( \barr{cc} {\mathbbm m}_1 & 0 \\ 0 & {\mathbbm m}_2 \earr \right)
\label{diag-1}
\eeqa
that diagonalize the mass matrix. From Eq.~(\ref{diag-1}) one gets
\beqa
{\mathbbm m}_1 &=& 
A^\dagger {\mathbbm m}_D C^\ast + C^\dagger {\mathbbm m}_D^T A^\ast + C^\dagger {\mathbbm M}_{\rm N} C^\ast \; , 
\label{m1}\\
{\mathbbm m}_2 &=& 
B^\dagger {\mathbbm m}_D D^\ast + D^\dagger {\mathbbm m}_D^T B^\ast + D^\dagger {\mathbbm M}_{\rm N} D^\ast \;.
\label{m2}
\eeqa
The relation in Eq.~(\ref{m2}) gives
\beqa
D {\mathbbm m}_2 D^T
&=& \Delta_D + ({\mathbbm 1} - \epsilon){\mathbbm M}_{\rm N}({\mathbbm 1} - \epsilon^T) \; ,
\eeqa
where $\Delta_D \equiv D B^\dagger {\mathbbm m}_D ({\mathbbm 1} - \epsilon^T) 
+ ({\mathbbm 1} - \epsilon) {\mathbbm m}_D^T B^\ast D^T$. Here 
we have used the fact that ${\mathbbm M}_{\rm N} \gg {\mathbbm m}_D $ and defined 
$D D^\dagger = {\mathbbm 1} - \epsilon$, where 
$\epsilon = \left( {\mathbbm m}_D {\mathbbm M}_{\rm N}^{-1} \right)^n$, $n$ to be determined. 
The unitarity condition gives $B^\dagger B + D^\dagger D = {\mathbbm 1}$, which 
implies $B = {\cal O}(\epsilon^{1/2})$. The same holds for $C$. In a similar 
way one has from unitarity $A A^\dagger = {\mathbbm 1} - \epsilon$. 
Eq.~(\ref{diag-1}) also gives
\beqa
A^\dagger {\mathbbm m}_D D^\ast + C^\dagger {\mathbbm m}_D^T B^\ast + C^\dagger {\mathbbm M}_{\rm N} D^\ast = 0 \; ,
\eeqa
which reduces to 
\beqa
A^\dagger {\mathbbm m}_D + C^\dagger {\mathbbm M}_{\rm N} &=& C^\dagger {\mathbbm m}_D^T B^\ast {(D^\ast)}^{-1} \nn \\
&=& {\cal O}(\epsilon^{1/2}){\mathbbm m}_D^T {\cal O}(\epsilon^{1/2}) D^T ({\mathbbm 1} + \epsilon^T) \; .
\eeqa
Thus
\beqa
C^\dagger = {\cal O}(\epsilon) - A^\dagger {\mathbbm m}_D {\mathbbm M}_{\rm N}^{-1} \; ,
\eeqa
which in turn shows that $\epsilon \sim {\cal O}({({\mathbbm m}_D /{\mathbbm M}_{\rm N})}^{2})$. 
Hence $B \sim C \sim {\cal O}({\mathbbm m}_D /{\mathbbm M}_{\rm N})$
and $D {\mathbbm m}_2 D^T = {\mathbbm M}_{\rm N} +{\cal O}({\mathbbm m}_D^2 /{\mathbbm M}_{\rm N}) $, 
which is the same as that given in Eq.~(\ref{seesaw-heavy}) in Sec~\ref{sec:mass-generation}.
From Eq.~(\ref{m1}), keeping terms upto ${\cal O}({\mathbbm m}_D /{\mathbbm M}_{\rm N})$, one 
gets
\beqa
A {\mathbbm m}_1 A^T = -{\mathbbm m}_D {\mathbbm M}_{\rm N}^{-1} {\mathbbm m}_D^T \; ,
\eeqa
and this is nothing but the seesaw relation, quoted in Eq.~(\ref{seesaw}). 
Related discussions can also be found in \cite{zing_review}.



\end{document}